\documentclass[11pt,a4paper]{article}

\usepackage{mathtools}
\usepackage{microtype}
\usepackage{authblk} 
\usepackage{algpseudocode,algorithmicx,algorithm}

\usepackage{mathrsfs}
\usepackage{latexsym,bm,bbm}
\usepackage{amsmath,amsfonts,amssymb,amsthm,rotating}
\usepackage{extarrows}

\usepackage{graphicx,subfigure,epstopdf,float}
\usepackage{placeins}
\graphicspath{{images/}{./}}
\usepackage{enumerate,cases,multirow}
\usepackage{makecell}
\usepackage{caption}

\usepackage{longtable,colortbl,arydshln,threeparttable}
\definecolor{mygray}{gray}{.9}

\usepackage{indentfirst}
\usepackage[top=25mm,bottom=20mm,left=25mm,right=20mm]{geometry}
\usepackage{cite}

\usepackage{makeidx}        
\usepackage{booktabs}
\usepackage{tikz}
\usepackage{xcolor}
\usepackage[bookmarksnumbered,colorlinks=true,citecolor=red,linkcolor=red,hyperindex,linktocpage=true]{hyperref}

\newcommand{\ket}[1]{| #1 \rangle} 
\newcommand{\bra}[1]{\langle #1 |} 

\newcommand{\bb}{\boldsymbol}

\DeclareMathOperator{\diag}{diag}

\newcounter{parentalgorithm}

\makeatother

\newtheorem{theorem}{Theorem}[section]
\newtheorem{lemma}{Lemma}[section]
\newtheorem{proposition}{Proposition}[section]

\theoremstyle{remark}
\newtheorem{remark}{\bf Remark}[section]

\numberwithin{equation}{section}

\hypersetup{pdftitle={A hybrid quantum-classical lattice kinetic scheme with tensor-train feedback for incompressible flows},pdfauthor={Yongwei Guo; Yuan Yu; Yue Yu; Yuting Zhou},pdfkeywords={Quantum algorithm; Lattice kinetic scheme; Incompressible Navier-Stokes equations; Hybrid quantum-classical algorithm; Tensor train; Block-encoding}}

\begin{document}

\title{\bfseries A hybrid quantum--classical lattice kinetic scheme with tensor-train feedback for incompressible flows}
\author[1]{Yongwei Guo}

\author[1,2,3]{Yuan Yu%
  \thanks{Corresponding author.
    \textit{Email addresses:}
    \texttt{yuyuan@xtu.edu.cn} (Yuan Yu),
    \texttt{terenceyuyue@xtu.edu.cn} (Yue Yu),
    \texttt{202521511280@smail.xtu.edu.cn} (Yuting Zhou)}}

\author[1,2,3]{Yue Yu}

\author[1]{Yuting Zhou}

\affil[1]{School of Mathematics and Computational Science,
  Xiangtan University, Xiangtan, 411105, Hunan, China}

\affil[2]{Hunan Research Center of the Basic Discipline Fundamental
  Algorithmic Theory and Novel Computational Methods,
  Xiangtan, 411105, Hunan, China}

\affil[3]{National Center for Applied Mathematics in Hunan,
  Xiangtan, 411105, Hunan, China}
\date{}

\maketitle

\begin{abstract}
Nonlinear lattice kinetic simulation requires each updated distribution to supply the velocity and gradients that determine the next step. In a quantum--classical solver, reading out the full distribution incurs work proportional to the lattice size. We construct a tensor-train (TT) interface that retains reusable classical cores at the physical amplitude scale for a full-relaxation lattice kinetic scheme. The cores supply nonlinear coefficients and prepare each quantum input. Energy-scaled propagation and coordinate recovery encode the updated distribution, and alternating least-squares reconstruction converts signed Hadamard-test overlaps into output cores by restoring a known normalization factor. The next step starts from accepted cores without replaying previous quantum evolution. Under positive, unit-sum reference coefficients, positive lower and finite upper density bounds, and Lipschitz equilibrium sensitivity, perturbation bounds relate local defects to feedback errors. Resource bounds account for classical operations, circuit execution, sampling, and verification. Classical diagnostics expose why distribution accuracy alone is insufficient: in 80-step feedback on a $16\times16$ grid, rank cap 4 nearly eliminates the vortex despite a final distribution error below $1\%$, whereas rank cap 12 gives relative velocity error $1.03\times10^{-3}$. Four-step binomial readout simulations show that tightening core sampling tolerance threefold raises the prescribed sufficient budget from $5.97\times10^{12}$ to $5.37\times10^{13}$ shots as the result approaches the fixed-rank, fixed-sweep exact-overlap trajectory. The interface therefore provides a way to connect quantum propagation to nonlinear feedback and to assess the separate demands of compression, velocity accuracy, and readout precision.
\end{abstract}


\section{Introduction}\label{sec:introduction}

Nonlinear flow simulation requires each updated state to supply the fields that determine the next time step. In lattice Boltzmann methods (LBM), local collision and lattice streaming evolve discrete distributions, while density and momentum moments recover the macroscopic flow variables \cite{hou1995simulation,sterling1996stability,kruger2017lattice,astoul2020analysis,renard2021linear}. The full-relaxation lattice kinetic scheme (LKS) of Xiao et al.~\cite{xiao2025latticekinetic} makes this dependence explicit: at unit dimensionless relaxation time ($\tau^*=1$), collision replaces the incoming distribution by a modified equilibrium determined by velocity and its symmetric spatial gradient. A scalar parameter $A_{\mathrm{LKS}}$ sets the nominal kinematic viscosity. Freezing these fields makes one collision--streaming step linear, but continuing the flow calculation requires recovering velocity and its gradient from the updated distribution. A compressed solver must therefore preserve information that can be reused in this nonlinear feedback.

Quantum algorithms have established several ways to propagate discretized differential equations and kinetic transport in compact state encodings \cite{berry2014high,todorova2020quantum,budinski2021quantum,Costa2021QLSA,jin2022quantum,Lin2023timeODE,schalkers2024efficient,jin2026schrodingerization,novikau2026explicit}. For prescribed-coefficient advection--diffusion, recent lattice constructions develop unitary or block-encoded collision and streaming, dynamic circuits, and multiple time steps without reinitialization \cite{wawrzyniak2024unitary,xu2025improved,wawrzyniak2025linearized,nagel2026multiple,georgescu2025fully}. Kumar and Frankel~\cite{kumar2025quantum} give a unitary matrix representation and low-Reynolds-number flow tests for a linearized equilibrium. These studies provide concrete propagation mechanisms and clarify their encoding costs. Our companion study~\cite{HeYu2026ADE} uses equilibrium energy scaling to incorporate zeroth-moment recovery into an orthogonal-projection collision factor under positive, unit-sum coefficients. For self-consistent flow, the propagation mechanism must be combined with a way to determine the coefficients of subsequent updates from the evolving distribution.

Nonlinear quantum simulation has also progressed through several distinct representations. Carleman, Liouville, and Koopman--von Neumann embeddings make linear quantum evolution applicable to nonlinear equations, with analyses that identify how truncation and data access enter their complexity \cite{liu2021efficient,jin2023time,schalkers2024importance}. Asztalos et al.~\cite{asztalos2024reduced} instead combine data-driven dynamic mode decomposition with quantum optimization to advance reduced-order models. In lattice fluid dynamics, Wang et al.~\cite{wang2025nonlinear} introduce a node-level ensemble description with transport, collision, and an H-step that restores a near-equilibrium ensemble structure; their classical simulations include vortex merging and decaying turbulence. Duong et al.~\cite{duong2026denoising} develop a geometric collision operator by projecting onto the linearized equilibrium manifold around a reference state, and investigate a route to coherent multiple-step execution. These approaches expand the available treatments of nonlinear flow. Their representations and approximation conditions differ from the full-relaxation LKS recurrence considered here, for which we seek reusable classical data that retain its moment-dependent coefficients.

Progress in state preparation and readout provides a complementary starting point. Kocherla et al.~\cite{kocherla2024fully} use iterative phase estimation for phase correction to reduce full-distribution measurement and re-encoding between quantum lattice steps. Tiwari et al.~\cite{tiwari2025algorithmic} combine matrix product state (MPS) loading with observable-based reconstruction, demonstrating repeated advection--diffusion steps on hardware for a Gaussian density parameterized by its mean vector and covariance matrix. This demonstrates how a suitable low-dimensional output description can support subsequent state preparation. For full-relaxation LKS, the required outputs are spatially resolved momentum-to-density ratios and velocity gradients. Without an assumed low-dimensional parameterization that determines these fields, a few scalar observables do not suffice. Reading out and reloading all $qN$ distribution entries, where $q$ is the number of discrete velocities and $N$ the number of nodes, would incur $\Omega(qN)$ classical work and storage per step. The task is therefore to reconstruct a representation that supplies these fields and retains the physical amplitude scale, with its readout and reloading costs included \cite{schalkers2024efficient,georgescu2025fully,novikau2026explicit}.

Tensor trains (TT), equivalently MPS, offer such a reusable classical representation \cite{schon2005sequential,oseledets2011TT,gelss2017nearest}. Low-rank tensor methods already evaluate moments and self-consistent fields on compressed factors for kinetic equations \cite{ehrlacher2017dynamical,boelens2018parallel,boelens2020tensor,guo2022low,einkemmer2025review,asgrim2026tensor} and nonlinear partial differential equations \cite{gorodetsky2018gradient,dektor2021dynamic,danis2025tensor,gidi2025pseudospectral}. Recent developments clarify how to preserve relevant physics and evaluate nonlinear terms: Coughlin et al.~\cite{coughlin2024robust} separate the macroscopic component carrying conserved moments from a low-rank microscopic component; Dektor and Einkemmer~\cite{dektor2026interpolatory} use interpolatory low-rank evolution for the 3+3-dimensional Boltzmann--BGK equation; and Wang and Hu~\cite{wang2026dynamical} advance velocity-space TT factors by projector splitting, treating position as a parameter. These results motivate compressed feedback and moment-aware accuracy checks, while the ranks needed for our spatial tensorization must be assessed separately. With binary spatial tensorization into $p=\mathcal{O}(\log N)$ modes, maximum mode size $n_{\max}$, and rank bound $r$, TT cores store $\mathcal{O}(pn_{\max}r^2)$ parameters. Orthogonalized cores also admit sequential isometric quantum state preparation \cite{schon2005sequential}. Thus classical TT cores can supply nonlinear coefficients and quantum inputs within one interface, provided that the required ranks remain manageable.

Returning the quantum output to this representation requires more than a small parameter count. The propagation circuit acts on an energy-scaled, normalized state, whereas the next classical step uses the distribution at its physical amplitude scale. Core reconstruction must connect these representations and preserve the moments that drive feedback. At low Mach number, the distribution norm is dominated by the density background, while velocity depends on much smaller differences between opposite-direction populations. A small distribution error can therefore coexist with an inaccurate velocity field. In addition, estimating each core entry from quantum overlaps introduces a sampling cost that storage compression alone does not control.

We construct a hybrid LKS interface that stores the unnormalized physical-scale distribution in classical TT cores between time steps. Classical contractions, a TT-rounded iteration for reciprocal density, and derivative matrix product operators (MPOs) form the velocity, gradients, and equilibrium coefficients. The quantum step uses the current coefficients for energy scaling and collision--streaming, then returns to physical coordinates within the circuit. Its output remains divided by a known overall normalization factor. Alternating least-squares (ALS) updates in orthogonalized TT frames estimate signed Hadamard-test overlaps and multiply back this factor to reconstruct physical-scale cores. Once the fitting residual and admissibility checks pass, these stored cores supply the next nonlinear update. Repeated readout uses fresh executions of the current step from its stored input, without replaying previous quantum steps.

The analysis identifies the conditions under which this interface controls feedback error and the resources needed to meet those conditions. The single-step construction retains the actual nodewise coefficient sum, including its discrete velocity-divergence correction. Under positive, unit-sum reference coefficients and the stated density and Lipschitz sensitivity assumptions, Proposition~\ref{prop:TTnonlinear} propagates local coefficient and update defects; the associated moment estimate explains the $\mathcal{O}(c/U)$ sensitivity of relative velocity error at low flow speed $U$ compared with lattice speed $c$. The work bounds in Section~\ref{sec:resources} account jointly for coefficient evaluation, circuit preparation, repeated overlap measurements, and residual checks. This accounting distinguishes the reduction in stored parameters from the precision-dependent cost of reconstructing them and supplies a basis for comparison with purely classical TT propagation at matched output tolerances.

The numerical studies test these requirements separately. Classical Taylor--Green refinement checks the discretization and scaled-update algebra, and a classical three-dimensional cavity snapshot tests static quantized TT compression. Dynamic feedback tests then examine whether compression preserves the evolving flow. In an 80-step exact-overlap test, rank cap 4 gives a final distribution error below $1\%$ while nearly eliminating the vortex; rank cap 12 reduces the relative velocity error against the uncompressed discrete solution to $1.03\times10^{-3}$. Four-step binomial simulations of overlap readout show how tighter sampling tolerances move the result toward the fixed-rank, fixed-sweep exact-overlap trajectory at an inverse-square sampling cost. These tests use classical reference operations and simulated measurement statistics to distinguish spatial compressibility, feedback accuracy, and readout precision.

Section~\ref{sec:LKSextension} defines the discrete flow model and required feedback outputs. Section~\ref{sec:TThybrid} constructs the TT interface, and Sections~\ref{sec:hybridAnalysis} and~\ref{sec:resources} analyze its error and resource requirements. Section~\ref{sec:numerics} presents the numerical diagnostics, followed by conclusions in Section~\ref{sec:conclusion}. Appendices~\ref{app:linearBE}--\ref{app:readoutBudget} give prescribed-coefficient propagation results, TT algebra, and a worked sampling budget.

\section{Discrete lattice kinetic model and nonlinear feedback requirements}\label{sec:LKSextension}

The output of one time step must contain enough information to define the next nonlinear operator. We make this requirement precise by specifying the full-relaxation LKS recurrence and separate accuracy criteria for the distribution and macroscopic feedback. The quantum construction and perturbation analysis use periodic streaming; the wall-bounded cavity calculation in Section~\ref{subsec:cavity3d} provides a separate classical discretization and static-compression test.

\subsection{Flow variables and the full-relaxation discretization}\label{subsec:flowModel}

The discrete kinetic carrier of our hybrid interface is the full-relaxation discretization of viscous incompressible flow, whose constant-density continuum equations on a periodic domain are
\[
 \partial_t\bb u+(\bb u\cdot\nabla)\bb u
 =-\varrho_{\mathrm{ref}}^{-1}\nabla p+\nu\Delta\bb u,
 \qquad \nabla\cdot\bb u=0.
\]
Here $\varrho_{\mathrm{ref}}$ is the reference density, $\nu$ the kinematic viscosity, and $\bb u$ and $p$ the velocity and pressure fields. The hybrid algorithm advances the discrete LKS recurrence defined next, which retains a variable kinetic density $\varrho^k$. Its approximation error relative to this recurrence is distinct from the consistency error relative to the continuum equations.

We discretize space on a uniform periodic Cartesian grid with $N$ spatial nodes in $d=2$ or $3$ dimensions, grid spacing $\Delta x$, time step $\Delta t$, and lattice speed $c=\Delta x/\Delta t$. The discrete velocities, quadrature weights, and lattice sound speed of the D2Q9 or D3Q27 stencil satisfy
\[
 \bb c_i=c\bb e_i,\qquad \bb e_i\in\{-1,0,1\}^d,
 \qquad w_i=\prod_{\gamma=1}^d\omega(e_{i,\gamma}),
 \quad \omega(0)=\frac23,\quad\omega(\pm1)=\frac16,
 \quad c_s^2=\frac{c^2}{3}.
\]
Thus the stencil contains $q=3^d$ discrete velocities, and the direction-summation matrix $E_I=[I_N,\ldots,I_N]$ formed from spatial identity blocks has $q$ blocks.

We denote the kinetic density by $\varrho$ to distinguish it from the weighted-streaming norm bound $\rho^k$. From the discrete distribution components at each lattice node and time level, the macroscopic density, momentum, and velocity are defined by
\begin{equation}\label{eq:LKSflowMoments}
 \varrho^k=\sum_{i=0}^{q-1}f_i^k,\qquad
 \bb m^k=\sum_{i=0}^{q-1}\bb c_i f_i^k,\qquad
 \bb u^k=\bb m^k/\varrho^k.
\end{equation}
For definiteness, we approximate the spatial velocity gradient by periodic centered differences
\begin{equation}\label{eq:LKSdiscreteGradient}
 (\nabla_h\bb u)_{\alpha\beta}(\bb x_j)
 =\frac{u_\alpha(\bb x_j+\Delta x\bb e_\beta)
             -u_\alpha(\bb x_j-\Delta x\bb e_\beta)}{2\Delta x},
 \qquad \alpha,\beta=1,\ldots,d,
\end{equation}
where $\bb e_\beta$ here denotes a Cartesian coordinate unit vector.

In the notation of this paper, the modified equilibrium given by formula~(19) of Xiao et al.~\cite{xiao2025latticekinetic} takes the form
\begin{equation}\label{eq:LKSequilibriumExplicit}
 \begin{aligned}
 f_i^{\mathrm{eq},k}(\bb x_j)&=\varrho^k(\bb x_j)a_i[\bb u^k](\bb x_j),\\
 a_i[\bb u^k](\bb x_j)&=w_i\left[1+
 \frac{\bb c_i\cdot\bb u^k(\bb x_j)}{c_s^2}
 +\frac{(\bb c_i\cdot\bb u^k(\bb x_j))^2}{2c_s^4}
 -\frac{|\bb u^k(\bb x_j)|^2}{2c_s^2}\right.\\
 &\hspace{3em}\left.{}+A_{\mathrm{LKS}}\Delta t\,
 \bb c_i^\top\bigl(\nabla_h\bb u^k(\bb x_j)
            +(\nabla_h\bb u^k(\bb x_j))^\top\bigr)\bb c_i\right].
 \end{aligned}
\end{equation}
Here $A_{\mathrm{LKS}}$ is a prescribed scalar parameter, denoted by $A$ in that study and distinct from the diagonal coefficient matrices $A_i^k$ and $\mathcal A^k$ used in our stacked operator notation. At full relaxation, its nominal kinematic viscosity relation reads
\begin{equation}\label{eq:LKSviscosity}
 \nu=\left(\frac12-2A_{\mathrm{LKS}}c_s^2\right)c_s^2\Delta t,
 \qquad
 A_{\mathrm{LKS}}=\frac{1}{2c_s^2}
       \left(\frac12-\frac{\nu}{c_s^2\Delta t}\right),
\end{equation}
matching formula~(20) of Xiao et al.~\cite{xiao2025latticekinetic} because the symmetric velocity gradient in~\eqref{eq:LKSequilibriumExplicit} omits a factor of $1/2$. Adopting this nominal viscosity relation identifies the target discrete LKS rather than establishing an error analysis of its incompressible limit.

In $a_i[\bb u](\bb x)$, brackets indicate dependence on the velocity
field, including its gradient through \eqref{eq:LKSdiscreteGradient};
parentheses specify the spatial point. At full relaxation ($\tau^*=1$), collision followed by periodic streaming advances the distribution according to
\begin{equation}\label{eq:LKSnonlinear}
 f_i^{k+1}(\bb x+\bb c_i\Delta t)
 =\varrho^k(\bb x)\,a_i[\bb u^k](\bb x),
 \qquad
 \varrho^k=\sum_i f_i^k,\quad
 \varrho^k\bb u^k=\sum_i\bb c_i f_i^k.
\end{equation}
Starting from an initial distribution $\bb f^0$ with positive density, each periodic step must recompute the macroscopic moments, velocity gradients, and equilibrium weights because the quadratic velocity terms, gradient correction, and momentum-to-density ratio render~\eqref{eq:LKSnonlinear} nonlinear in the evolving state.

\subsection{Frozen-coefficient map and required feedback outputs}\label{subsec:frozenFeedback}

To separate the linear operator executed on the quantum circuit from the nonlinear state information required across time steps, we rewrite one full-relaxation update in stacked matrix form. Let $N$ be the number of spatial nodes and stack $\bb f^k=(\bb f_0^k,\ldots,\bb f_{q-1}^k)^\top$ by direction. Periodic streaming is the permutation $P=\diag(P_0,\ldots,P_{q-1})$, with $P_i\ket j=\ket{\pi_i(j)}$. The matrix $E_I=[I_N,\ldots,I_N]$ sums direction blocks.

Let $\bb C$ be the moment operator whose block for direction $i$
is $\bb c_i\otimes I_N$, with momentum components stacked consistently.
For any distribution vector $\bb v$ with positive nodal density, we first evaluate the macroscopic fields
\begin{equation}\label{eq:LKSfeedbackMap}
 \begin{aligned}
  \bb\varrho(\bb v)&=E_I\bb v,& \bb m(\bb v)&=\bb C \bb v,&
  \bb u(\bb v)&=\bb m(\bb v)/\bb\varrho(\bb v).
 \end{aligned}
\end{equation}
Here division acts pointwise across lattice nodes using the same nodal density for each momentum component. Stacking the equilibrium weights in the same direction--space ordering as the distribution via
\[
 \bb a[\bb u]_{i,j}=a_i[\bb u](\bb x_j),
\]
and evaluating this vector at the moment-derived velocity yields $\bb a[\bb u(\bb v)]$, so the nonlinear step operator takes the form
\[
 \Phi(\bb v)=P\diag\bigl(\bb a[\bb u(\bb v)]\bigr)E_I^\top E_I\bb v.
\]
The full-relaxation LKS is $\bb f^{k+1}=\Phi(\bb f^k)$. Evaluating the moments at $\bb v=\bb f^k$ and freezing the resulting equilibrium coefficients within step $k$ gives the frozen-coefficient representation
\begin{equation}\label{eq:LKSvelocityDependentMatrix}
 \bb f^{k+1}=M^k\bb f^k,\qquad
 M^k=P\mathcal A^kE_I^\top E_I,\qquad
 \mathcal A^k=\diag_{i,j}\bigl(a_i[\bb u^k](\bb x_j)\bigr),
 \quad \bb u^k=\frac{\bb C\bb f^k}{E_I\bb f^k}.
\end{equation}

Although $M^k$ acts linearly once $\mathcal A^k$ is fixed, advancing to step $k+1$ requires reusable classical data from the updated state to compute the next density, velocity, and spatial gradient. A normalized quantum state alone does not meet this feedback requirement.

We denote the exact discrete distribution by $\bb f^k$ and its physical-scale TT approximation, incorporating compression, coefficient, and fitting errors, by $\bb f_{\mathrm{TT}}^k$, with corresponding computed velocity $\bb u_{\mathrm{TT}}^k$.

When comparing implementations, we prescribe separate tolerances for the physical distribution and for the macroscopic velocity and velocity-gradient feedback,
\begin{equation}\label{eq:LKSoutputRequirements}
 \begin{aligned}
 \|\bb f_{\mathrm{TT}}^{k}-\bb f^k\|&\leq\varepsilon_f,\\
 \|\bb u_{\mathrm{TT}}^{k}-\bb u^k\|
       +\Delta x\|\nabla_h\bb u_{\mathrm{TT}}^{k}
                         -\nabla_h\bb u^k\|&\leq\varepsilon_u.
 \end{aligned}
\end{equation}
Tracking these accuracy criteria separately prevents a small full-distribution error from masking inaccurate macroscopic velocity or gradient feedback.

\section{Hybrid time marching with tensor-train feedback}\label{sec:TThybrid}

The hybrid scheme meets the feedback requirements of Section~\ref{sec:LKSextension} by retaining a classical TT representation of the physical distribution after each accepted step. The stored cores provide the next coefficients and prepare the next quantum input. The remaining task is to convert the current quantum output back into cores at the correct physical scale. We describe this cycle in execution order: initialization, classical coefficient evaluation, quantum propagation and coordinate recovery, then core reconstruction and acceptance.

\subsection{Physical-scale TT state, initialization, and step interface}\label{subsec:TTworkingState}\label{subsec:TTinitialization}\label{subsec:LKSproblemStrategy}

Maintaining nonlinear feedback across steps requires an unnormalized classical representation from which local moments and spatial gradients can be contracted directly. We fix an index bijection $(i,j)\leftrightarrow(\xi_1,\ldots,\xi_p)$ with mode sizes $n_s$ satisfying $\prod_s n_s=qN$, placing the velocity direction first and tensorizing each spatial coordinate into binary digits whenever the grid admits a quantized TT (QTT) layout; this mode ordering remains fixed throughout a simulation. Storing classical cores $\mathsf F_s^k\in\mathbb R^{r_{s-1}^k\times n_s\times r_s^k}$ with boundary ranks $r_0^k=r_p^k=1$ represents the unnormalized physical distribution as
\begin{equation}\label{eq:TTrepresentation}
 (\bb f_{\mathrm{TT}}^k)_{i,j}
 =\mathsf F_1^k(\xi_1)\Join\mathsf F_2^k(\xi_2)\Join\cdots
                         \Join\mathsf F_p^k(\xi_p),
 \qquad \mathsf F_s^k(\xi_s)\in
          \mathbb R^{r_{s-1}^k\times r_s^k}.
\end{equation}

Here $\Join$ contracts adjacent auxiliary indices, so fixing all physical indices reduces the chain to a product of matrix slices and hence a scalar. Because the TT ranks $r_s^k$ govern compression according to the state structure and target accuracy, the cores store $\sum_s n_sr_{s-1}^kr_s^k\leq pn_{\max}r^2$ entries whenever $n_s\leq n_{\max}$ and $r_s^k\leq r$.

Computed moments, velocity, and coefficients carry the subscript $\mathrm{TT}$. Coefficient cores $\mathsf A_s^k$ use the same index mapping but may have different ranks; they are distinct from the diagonal operator $\mathcal A_{\mathrm{TT}}^k=\diag(\bb a_{\mathrm{TT}}^k)$. Linear operations such as discrete shifts and derivatives are represented as TT matrices, also called MPOs, through their matrix cores. The symbol $\odot$ denotes pointwise multiplication. Rounding $\mathcal R_\delta$ denotes compression with certified absolute Euclidean error at most $\delta$; a rank cap alone does not certify this tolerance. Spatial coordinate tensorization, MPO arithmetic, and TT rounding rules for these operations are detailed in Appendix~\ref{subsec:TTrepresentation}.

At $k=0$, suppose first that the classical initial values $f_{i,j}^0$ are supplied. Using the fixed index mapping, we reshape this array into an $n_1\times\cdots\times n_p$ tensor and apply TT-SVD successively~\cite{oseledets2011TT}: each step reshapes the remaining tensor into a matrix with $r_{s-1}^0n_s$ rows, retains the selected left singular vectors as core $\mathsf F_s^0$, and passes the singular values multiplied by the right singular vectors to the subsequent factorization until the final remainder forms the last core. The resulting cores define $\bb f_{\mathrm{TT}}^0$ through~\eqref{eq:TTrepresentation}, with singular-value truncations chosen to satisfy an initial error tolerance $\|\bb f_{\mathrm{TT}}^0-\bb f^0\|\leq\delta_0$ via the classical preprocessing map
\[
 \{f_{i,j}^0\}_{i,j}
 \xrightarrow{\text{reshape and TT-SVD}}\{\mathsf F_s^0\}_{s=1}^p
 \quad\Longrightarrow\quad \bb f_{\mathrm{TT}}^0\approx\bb f^0.
\]
Whereas full-array initialization reads all $qN$ entries, analytically structured initial data may admit direct core construction: a separable product $f^0(\xi_1,\ldots,\xi_p)=\prod_s b_s(\xi_s)$ yields rank-one cores $\mathsf F_s^0(\xi_s)=b_s(\xi_s)$, while finite sums admit explicit cores followed by TT rounding. In any initialization route, a normalized quantum initial state alone cannot supply these classical cores.

At each time step $k\geq0$, given the physical-scale input cores $\{\mathsf F_s^k\}$, the classical stage computes density and momentum, a rounded reciprocal density, velocity gradients, and the coefficients in \eqref{eq:LKSequilibriumExplicit}. These coefficients remain fixed during the quantum step and all subsequent fitting sweeps, defining the frozen physical target
\begin{equation}\label{eq:TToverviewTarget}
 \bb y^k=M_{\mathrm{TT}}^k\bb f_{\mathrm{TT}}^k,
 \qquad M_{\mathrm{TT}}^k
 =P\diag(\bb a_{\mathrm{TT}}^k)E_I^\top E_I.
\end{equation}
The target $\bb y^k$ fixes what the output cores must approximate. Section~\ref{subsec:TTcoefficients} first constructs its coefficient cores $\{\mathsf A_s^k\}$. Section~\ref{subsec:TTquantumStep} then encodes $\bb y^k$ in quantum amplitudes divided by a known normalization factor. Section~\ref{subsec:TTreadout} uses signed overlaps and this factor to reconstruct and check $\{\mathsf F_s^{k+1}\}$, which become the next classical input.

\subsection{Classical evaluation of the nonlinear coefficients}
\label{subsec:TTcoefficients}

To form the frozen operator $M_{\mathrm{TT}}^k$ in \eqref{eq:TToverviewTarget} without expanding onto the $qN$ lattice, all macroscopic moments, density reciprocals, spatial gradients, and equilibrium weights are evaluated directly on classical TT cores. Starting from $\bb f_{\mathrm{TT}}^{k}$, we first contract the direction index to form
 $\bb\varrho_{\mathrm{TT}}^k=E_I\bb f_{\mathrm{TT}}^{k}$ and
 $\bb m_{\mathrm{TT}}^{k}=\bb C\bb f_{\mathrm{TT}}^{k}$.
 These moment evaluations are classical TT contractions, with any subsequent rounding recorded in the step error budget.

Next, we evaluate the pointwise reciprocal of the density in classical TT format to obtain a vector $\bb d$ satisfying $d_j\approx1/\varrho_{\mathrm{TT},j}^k$, enabling momentum-to-velocity conversion by pointwise multiplication. Because the density and physical time level $k$ remain fixed throughout this inner loop, the superscript $\ell$ counts reciprocal iterations rather than physical time steps. Given certified bounds $0<\varrho_{\min}\leq\varrho_{\mathrm{TT},j}^k\leq\varrho_{\max}$, we iterate the TT-rounded Newton update
\begin{equation}\label{eq:TTreciprocal}
 \bb d^{(\ell+1)}=\mathcal R_{\delta^{\mathrm{inv}}_\ell}
 \bigl(\bb d^{(\ell)}\odot(2\bb 1-\bb\varrho_{\mathrm{TT}}^k\odot \bb d^{(\ell)})\bigr)
\end{equation}
Here $\odot$ denotes pointwise multiplication, $\bb 1$ is the all-ones vector, and $\mathcal R_{\delta^{\mathrm{inv}}_\ell}$ compresses the updated tensor with certified absolute Euclidean error at most $\delta^{\mathrm{inv}}_\ell$. We initialize this iteration with $\bb d^{(0)}=\bb 1/\varrho_{\max}$; without rounding, the nodal update reduces to
 \[
 d_j^{(\ell+1)}=d_j^{(\ell)}
       \bigl(2-\varrho_{\mathrm{TT},j}^k d_j^{(\ell)}\bigr).
 \]
For the scalar residual $r_j^{(\ell)}=1-\varrho_{\mathrm{TT},j}^k d_j^{(\ell)}$, this nodal recurrence gives $r_j^{(\ell+1)}=(r_j^{(\ell)})^2$.
 The stated initialization ensures
 $0\leq r_j^{(0)}\leq1-\varrho_{\min}/\varrho_{\max}<1$, so the
 unrounded iterates converge to the pointwise reciprocal.

 All operations act on classical cores. Rounding controls rank growth
 but perturbs the residual-squaring identity, so the residual must be
 checked. In particular,
 $\|\bb 1-\bb\varrho_{\mathrm{TT}}^k\odot\bb d\|_\infty\leq\eta$
 gives $\|\bb d-1/\bb\varrho_{\mathrm{TT}}^k\|_\infty
 \leq\eta/\varrho_{\min}$, where division is pointwise.
 A certified Euclidean residual bound also bounds the maximum norm.

Finally, we form the velocity approximation $\bb u_{\mathrm{TT}}^k\approx\bb m_{\mathrm{TT}}^k\odot\bb d$, apply derivative MPOs to obtain $\nabla_h\bb u_{\mathrm{TT}}^k$, and evaluate the equilibrium polynomial~\eqref{eq:LKSequilibriumExplicit} in TT format while recording all arithmetic and truncation errors within $\|\bb a_{\mathrm{TT}}^k-\bb a[\bb u(\bb f_{\mathrm{TT}}^k)]\|\leq\epsilon_a^k$.

Entrywise density bounds and coefficient admissibility conditions must be verified together with certified approximation tolerances, because sampling a few lattice nodes cannot establish global positivity. Although pointwise products, reciprocal iterations, and discrete gradient evaluations may increase intermediate TT ranks, all such operations act purely on stored classical cores and are counted in the classical work budget.

\subsection{Current-weight quantum propagation and physical amplitude recovery}
\label{subsec:TTquantumStep}

Once the classical stage supplies the distribution cores $\{\mathsf F_s^k\}$ and frozen coefficient cores $\{\mathsf A_s^k\}$ at step $k$, the quantum stage must apply the linear map $M_{\mathrm{TT}}^k$ even though $M_{\mathrm{TT}}^k$ need not be contractive in the Euclidean norm. We therefore propagate in current-weight coordinates induced by $\bb a_{\mathrm{TT}}^k$ and recover the physical-scale amplitudes inside the circuit before core fitting. Assume certified bounds
$0<a_{\min}\leq a_{i,\mathrm{TT}}^k(\bb x_j)\leq a_{\max}$,
including coefficient errors. The identities below describe the ideal
circuit; finite-precision errors enter the accepted step residual.

To prepare the weighted quantum input, define the diagonal operator and the scaled distribution
\begin{equation}\label{eq:TThybridPhysicalBlock}
 \mathcal A_{\mathrm{TT}}^k=\diag(\bb a_{\mathrm{TT}}^k),\qquad
 M_{\mathrm{TT}}^k=P\mathcal A_{\mathrm{TT}}^kE_I^\top E_I,
 \qquad
 \bb g_{\mathrm{TT}}^k=(\mathcal A_{\mathrm{TT}}^k)^{-1/2}
                                      \bb f_{\mathrm{TT}}^k.
\end{equation}
The matrix $\mathcal A_{\mathrm{TT}}^k$ is distinct from the coefficient cores $\mathsf A_s^k$. We approximate the pointwise inverse square root of the coefficient field, for example by evaluating a polynomial approximation on $[a_{\min},a_{\max}]$ via TT products and rounding, and multiply that factor into the distribution in classical TT arithmetic under a separate error budget. This scaled distribution may have TT ranks different from the physical one, and we contract its cores to obtain the normalization factor $n_{g,k}=\|\bb g_{\mathrm{TT}}^k\|$ and retain this norm classically. Writing $n_{\mathrm{sys}}$ for the number of system qubits in the qubit encoding of the TT modes (including any mode padding), we orthogonalize the scaled cores and extend their isometries to unitaries so as to prepare
\[
 U_{\mathrm{in}}\ket{0^{n_{\mathrm{sys}}}}=\ket{g_{\mathrm{TT}}^k}
 =\frac{\bb g_{\mathrm{TT}}^k}{n_{g,k}}.
\]
Sequential TT state preparation uses a rank-dependent bond register that is disentangled upon completion~\cite{schon2005sequential}; suppressing clean work registers and fixing a consistent phase convention for controlled calls, we reuse the same circuit construction for readout test vectors while retaining the physical cores in classical storage and treating the scaled cores as temporary working data.

The quantum step also requires coherent access to the coefficients stored in classical TT cores. Given a computational basis index $(i,j)$, a reversible coefficient-evaluation circuit (value oracle) multiplies the TT core slices selected by the index digits to compute $a_{i,\mathrm{TT}}^k(\bb x_j)$ and supply the square roots, ratios, and rotation angles required by the weighted collision, streaming, and recovery block-encodings. Temporary arithmetic registers are uncomputed after use, and the oracle synthesis, arithmetic precision, and TT-rank overhead enter the data-loading cost.

With the scaled input state and reversible coefficient access in place, the single-step evolution conjugates the frozen update by the current weight on both sides, factoring into weighted streaming and symmetric collision operators according to
\begin{equation}\label{eq:TTcurrentWeightStep}
 \begin{aligned}
 \overline B_{\mathrm{TT}}^k
 &=(\mathcal A_{\mathrm{TT}}^k)^{-1/2}
         M_{\mathrm{TT}}^k(\mathcal A_{\mathrm{TT}}^k)^{1/2}
   =\overline P_{\mathrm{TT}}^k C_{\mathrm{TT}}^k,\\
 \overline P_{\mathrm{TT}}^k
 &=(\mathcal A_{\mathrm{TT}}^k)^{-1/2}
                  P(\mathcal A_{\mathrm{TT}}^k)^{1/2},\\
 C_{\mathrm{TT}}^k&=Q_{\mathrm{TT}}^k(Q_{\mathrm{TT}}^k)^\dagger,
 \qquad Q_{\mathrm{TT}}^k=(\mathcal A_{\mathrm{TT}}^k)^{1/2}E_I^\top.
 \end{aligned}
\end{equation}
If the computed coefficients sum to one at every node, then
$(Q_{\mathrm{TT}}^k)^\dagger Q_{\mathrm{TT}}^k=I$ and the collision
factor is an orthogonal projection with normalization one, exactly as
in the full-relaxation case of Appendix~\ref{subsec:structuredBlockEncoding}.
Positive coefficients that do not sum to one require an additional
factor. Set $s_j^k=\sum_i a_{i,\mathrm{TT}}^k(\bb x_j)$ and choose
known bounds
\[
 s_{\max}^k\geq\max_j s_j^k,\qquad
 \overline\rho^k\geq\max_{i,j}
 \sqrt{\frac{a_{i,\mathrm{TT}}^k(\bb x_j)}
             {a_{i,\mathrm{TT}}^k(\bb x_{\pi_i(j)})}},
 \qquad \overline\alpha_k=\overline\rho^k s_{\max}^k.
\]
At each node, preparing the normalized direction state with amplitudes $\sqrt{a_{i,\mathrm{TT}}^k/s_j^k}$ and scaling its orthogonal projection by the diagonal sum factor $s_j^k$ yields $C_{\mathrm{TT}}^k$ as a block-encoding with normalization $s_{\max}^k$, while weighted streaming uses the upstream-to-downstream square-root coefficient ratio with normalization $\overline\rho^k$. Composing these two encodings with independent flag registers produces the one-step QLBM unitary $U_{\overline B}^k$ on $a_B$ flag qubits satisfying
\begin{equation}\label{eq:TToneStepBlockEncoding}
 (\bra{0^{a_B}}\otimes I)U_{\overline B}^k(\ket{0^{a_B}}\otimes I)
       =\frac{\overline B_{\mathrm{TT}}^k}{\overline\alpha_k}.
\end{equation}
We set $s_{\max}^k=1$ only when the discrete nodewise coefficient sum equals one identically; continuum incompressibility alone does not guarantee this discrete identity for the gradient-corrected equilibrium because nonzero discrete velocity divergence perturbs the nodewise sum, as shown in Section~\ref{subsec:LKSenergyConditions}.

Core fitting requires the physical output, which we recover within the quantum circuit.
The scaled step produces the temporary vector
\[
 \overline{\bb g}_{\mathrm{TT}}^{\,k+1}
   =\overline B_{\mathrm{TT}}^k\bb g_{\mathrm{TT}}^k
   =(\mathcal A_{\mathrm{TT}}^k)^{-1/2}\bb y^k,
 \qquad \bb y^k=M_{\mathrm{TT}}^k\bb f_{\mathrm{TT}}^k.
\]
This intermediate vector remains scaled by the current weight rather than the still-unknown next-step weight. To undo that scaling inside the circuit, we apply a diagonal recovery block-encoding $U_{\mathrm{rec}}^k$ of $(\mathcal A_{\mathrm{TT}}^k)^{1/2}/\sqrt{a_{\max}}$ implemented via coefficient-controlled rotations, and compose it with the weighted step unitary using independent flags to form
\[
 U_{\mathrm{step}}^k=U_{\mathrm{rec}}^k U_{\overline B}^k,
 \qquad \alpha_{\mathrm{hyb}}^k=\sqrt{a_{\max}}\,\overline\alpha_k.
\]
With identity operators on untouched flag registers left implicit, the recovery rotation adds one flag qubit so that the combined flag register uses $a_{\mathrm{flag}}=a_B+1$ qubits, and acting on the prepared scaled input yields the successful component
\begin{equation}\label{eq:TTscaledPhysicalOutput}
 U_{\mathrm{step}}^k
       \bigl(\ket{0^{a_{\mathrm{flag}}}}_{\mathrm{flag}}\ket{g_{\mathrm{TT}}^k}\bigr)
 =\ket{0^{a_{\mathrm{flag}}}}_{\mathrm{flag}}\frac{\bb y^k}
                                  {\alpha_{\mathrm{hyb}}^k n_{g,k}}
       +\ket\perp,\qquad
 (\bra{0^{a_{\mathrm{flag}}}}_{\mathrm{flag}}\otimes I)\ket\perp=0.
\end{equation}
The recovery block returns the output to physical coordinates, but the quantum amplitudes still contain the overall divisor $\alpha_{\mathrm{hyb}}^k n_{g,k}$. This factor combines input normalization with the block-encoding normalization. It is known from the current-step construction and is multiplied back into the measured overlaps when the classical output cores are formed.

Using the current weight in this manner forms an exact algebraic change of coordinates within one step,
\[
 (\mathcal A_{\mathrm{TT}}^k)^{1/2}\overline B_{\mathrm{TT}}^k
 (\mathcal A_{\mathrm{TT}}^k)^{-1/2}\bb f_{\mathrm{TT}}^k
 =M_{\mathrm{TT}}^k\bb f_{\mathrm{TT}}^k.
\]
Because the next equilibrium weight is evaluated only after reconstructing the physical output, this local coordinate change does not replace the prescribed-coefficient two-time-level scaling in~\eqref{gtime}; the resulting inter-step weighted-norm conversion is analyzed in Section~\ref{subsec:TTweightChanges}.

\subsection{Core reconstruction, residual acceptance, and nonlinear feedback}
\label{subsec:TTreadout}\label{subsec:TTfeedback}

Because \eqref{eq:TTscaledPhysicalOutput} leaves the updated distribution encoded in quantum amplitudes rather than classical memory, closing the time step requires reconstructing classical TT cores for the physical output of Section~\ref{subsec:TTquantumStep} without full-state tomography. Fix the input and coefficients at step $k$, and write
\[
 \bb y=\bb y^k=M_{\mathrm{TT}}^k\bb f_{\mathrm{TT}}^{k},\qquad
 n_{g,k}=\|\bb g_{\mathrm{TT}}^{k}\|>0.
\]
Here $\bb y$ denotes the encoded target, not an available classical
array. We fit it directly using candidate cores $\mathsf V_s$,
initialized, for example, by $\mathsf V_s=\mathsf F_s^k$.
Holding the remaining cores fixed makes the candidate linear in core $s$, reducing its local least-squares update to inner products with the encoded target.

To update core $s$, we orthogonalize the other candidate cores so that their contraction defines an isometric frame $\mathsf X_s\in\mathbb R^{qN\times d_s}$ with $\mathsf X_s^\dagger\mathsf X_s=I$. Here $d_s=n_sr_{s-1}r_s$ counts the entries in the candidate core. The frame maps these entries to a full distribution while holding the remaining cores fixed, so the local least-squares problem reduces to
\begin{equation}\label{eq:TTmarchCoreFit}
 \min_{\bb\theta_s\in\mathbb R^{d_s}}\|\mathsf X_s\bb\theta_s-\bb y\|^2,
 \qquad \bb\theta_s=\mathsf X_s^\dagger \bb y.
\end{equation}
Here $\bb\theta_s=\operatorname{vec}(\mathsf V_s)$ lists the entries of the candidate core $\mathsf V_s$ in a fixed order, with $\mathsf V_s$ acting as the output fitting variable while $\mathsf F_s^k$ remains the frozen input core. Because the frame columns are orthonormal, this projection avoids any linear-system inversion. Moreover, each frame column $\bb x_{s,t}=\mathsf X_s \bb e_t$, $1\leq t\leq d_s$, is itself a normalized TT vector formed from the fixed orthogonalized cores and a standard basis core, so its preparation unitary $U_{\bb x_{s,t}}$ compiles directly via the sequential isometric TT state-preparation circuit.

Let $U_{\mathrm{in}}\ket{0^{n_{\mathrm{sys}}}}=\bb g_{\mathrm{TT}}^{k}/n_{g,k}$. Including all block-encoding flags, we define the composite overlap unitary
\[
 W_{s,t}=(I_{\mathrm{flag}}\otimes U_{\bb x_{s,t}}^\dagger)
          U_{\mathrm{step}}^k(I_{\mathrm{flag}}\otimes U_{\mathrm{in}}).
\]
Applying its controlled implementation in a Hadamard test accesses the all-zero matrix element
\begin{equation}\label{eq:TTcoreOverlapReadout}
 \begin{aligned}
 &(\bra{0^{a_{\mathrm{flag}}}}\otimes\bra{0^{n_{\mathrm{sys}}}})
 W_{s,t}
 (\ket{0^{a_{\mathrm{flag}}}}\otimes\ket{0^{n_{\mathrm{sys}}}})\\
 &\qquad=\frac{\langle \bb x_{s,t},\bb y\rangle}{\alpha_{\mathrm{hyb}}^k n_{g,k}}
 =\frac{(\bb\theta_s)_t}{\alpha_{\mathrm{hyb}}^k n_{g,k}}.
 \end{aligned}
\end{equation}
A real-part Hadamard test estimates this signed real overlap, while a second quadrature would handle complex-valued tensor extensions. Because every measurement shot repeats input preparation, the weighted propagation and recovery circuit, and the inverse test-state circuit, the per-entry circuit cost depends on both the time level and the fitted core.

We then multiply each overlap estimate by the known factor $\alpha_{\mathrm{hyb}}^k n_{g,k}$, reshape the resulting $d_s$ entries into the updated core, and shift the orthogonality center to the adjacent core; explicitly, for a measured real overlap estimate $\widehat\mu_{s,t}$, the classical update reads
\[
 (\bb\theta_s)_t\leftarrow\alpha_{\mathrm{hyb}}^k n_{g,k}\widehat\mu_{s,t},
 \qquad
 \mathsf V_s\leftarrow\operatorname{reshape}
          (\bb\theta_s;r_{s-1},n_s,r_s).
\]
Sweeping across all cores and recomputing the orthogonalized frames after each core update refines the candidate tensor while the step input and encoded target remain fixed, with every overlap measurement using fresh executions of the current-step circuit.

Once the alternating sweeps produce a candidate distribution $\bb v$ with cores $\{\mathsf V_s\}$, the known normalization $\alpha_{\mathrm{hyb}}^k n_{g,k}$ retains the physical scale without first estimating $\|\bb y\|$, and acceptance is governed by the fitting residual. We evaluate this residual through available TT contractions or the norm-and-overlap identity
\begin{equation}\label{eq:TTquantumFitResidual}
 \|\bb y-\bb v\|^2=\|\bb y\|^2+\|\bb v\|^2-2\operatorname{Re}\langle \bb v,\bb y\rangle.
\end{equation}
Here $\|\bb v\|$ is the norm of the candidate distribution represented by the fitted cores and is computed by contracting those cores. The squared target norm $\|\bb y\|^2$ is obtained from
the success probability of
$U_{\mathrm{step}}^k(\ket{0^{a_{\mathrm{flag}}}}_{\mathrm{flag}}
\otimes\ket{g_{\mathrm{TT}}^k})$ multiplied by
$(\alpha_{\mathrm{hyb}}^k n_{g,k})^2$, and the cross term is obtained from the same
overlap test with $\bb v/\|\bb v\|$ in place of $\bb x_{s,t}$, after restoring the known physical amplitude scale and multiplying by the candidate distribution norm. In this expression
$\ket{g_{\mathrm{TT}}^k}=\bb g_{\mathrm{TT}}^{k}/n_{g,k}$.
Measurement uncertainty and errors in preparing the scaled input,
implementing the step, and recovering the physical distribution must
be included in the stopping test.
If the residual exceeds tolerance, we increase the candidate ranks or overlap precision and repeat the sweep, since a fixed-rank ALS sweep carries no guaranteed convergence to a prescribed tolerance. Once the residual and admissibility checks have passed, we accept $\bb v$ as $\bb f_{\mathrm{TT}}^{k+1}$ by storing $\mathsf F_s^{k+1}=\mathsf V_s$ (after any final orthogonalization), so the accepted cores define the updated distribution approximation via
\[
 (\bb f_{\mathrm{TT}}^{k+1})_{i,j}
 :=\mathsf F_1^{k+1}(\xi_1)\Join\cdots\Join\mathsf F_p^{k+1}(\xi_p)
 \approx (M_{\mathrm{TT}}^k\bb f_{\mathrm{TT}}^k)_{i,j}.
\]
These accepted, stored core entries constitute the quantum-to-classical output of the step.

The accepted cores become the next classical input, closing the single-step feedback loop
\begin{equation}\label{eq:TTfeedbackWorkflow}
\begin{gathered}
 \{\mathsf F_s^k\}
 \xrightarrow{\text{classical moments and equilibrium}}\{\mathsf A_s^k\},\\
 (\{\mathsf F_s^k\},\{\mathsf A_s^k\})
 \xrightarrow{\text{circuit construction}}(U_{\mathrm{in}},U_{\mathrm{step}}^k),\\
 (U_{\mathrm{in}},U_{\mathrm{step}}^k)
 \xrightarrow{\text{overlap measurements}}\{\bb\theta_s\}
 \xrightarrow{\text{classical core fitting}}\{\mathsf F_s^{k+1}\}.
\end{gathered}
\end{equation}
The accepted cores provide the physical distribution needed for both parts of the next step: classical feedback evaluation and quantum input preparation. The acceptance checks include density positivity and moments as well as the fitting residual. Step $k+1$ recomputes the coefficients and energy scaling from $\{\mathsf F_s^{k+1}\}$, checks positivity, and holds these data fixed during its fitting sweeps. Every new overlap measurement starts from this stored input; the execution costs in Section~\ref{subsec:quantumExecCost} therefore involve repeated current-step circuits rather than repeated histories.

For comparison, the same frozen update can be evaluated purely on classical cores through directionwise TT multiplication, periodic shifting, and tolerance-certified rounding according to
\begin{equation}\label{eq:TTnonlinearMarch}
 \bb f_{i,\mathrm{TT}}^{k+1}
 =\mathcal R_{\delta_{i,k}}
      \bigl(P_i(\bb a_{i,\mathrm{TT}}^k\odot\bb\varrho_{\mathrm{TT}}^k)\bigr).
\end{equation}
Any resource comparison against this purely classical baseline must include the full hybrid work budget spanning coefficient evaluation, circuit synthesis, repeated overlap shots, and residual verification; classical statistical simulations of the readout interface and its shot-noise scaling are presented in Section~\ref{subsec:TTnumericalExample}.

\section{Admissibility and conditional error propagation}\label{sec:hybridAnalysis}

Reconstructing an output distribution closes the algorithmic cycle, but errors in that distribution also change the coefficients of the next step. We first identify the coefficient conditions needed to encode one update, then relate changes of weighted norm to physical errors and derive a conditional feedback bound. These two levels require different assumptions: the single-step circuit retains the sum factor for positive coefficients whose nodewise sum differs from one, whereas the multi-step propositions use positive, unit-sum reference coefficients. The numerical diagnostics in Section~\ref{sec:numerics} retain this distinction.

\subsection{Coefficient conditions for energy-scaled propagation}
\label{subsec:LKSenergyConditions}

To separate the coefficient conditions required by the single-step circuit from the stricter reference-trajectory hypotheses of the multi-step perturbation bounds, first prescribe a velocity trajectory $\overline{\bb u}^{\,k}$ and evaluate its coefficients using \eqref{eq:LKSequilibriumExplicit}.
The entire velocity trajectory is prescribed before the linear solve; its values may nevertheless vary in space and time. With
$\mathcal A^k=\diag_{i,j}(a_i[\overline{\bb u}^{\,k}](\bb x_j))$, the resulting prescribed-coefficient linear recurrence reads
\begin{equation}\label{eq:LKSfrozen}
 \bb f^{k+1}=M^k\bb f^k,\qquad
 M^k=P\mathcal A^kE_I^\top E_I.
\end{equation}
This prescribed-trajectory recurrence reproduces the nonlinear LKS only when the prescribed velocity equals the velocity recovered from the evolving distribution moments.

Two algebraic properties determine whether the normalized-coefficient block-encoding construction in Appendix~\ref{app:linearBE} transfers directly, namely
\begin{equation}\label{eq:LKScoefficientConditions}
 a_i[\overline{\bb u}^{\,k}](\bb x_j)\geq a_*>0,\qquad
 \sum_{i=0}^{q-1}a_i[\overline{\bb u}^{\,k}](\bb x_j)=1.
\end{equation}
These positivity and unit-sum conditions must hold for the actual discrete coefficients, including the velocity-gradient correction. In particular, a continuum incompressibility condition does not automatically imply an exact discrete moment identity.

For the explicit coefficients in \eqref{eq:LKSequilibriumExplicit},
the identities $\sum_iw_i=1$, $\sum_iw_i\bb c_i=0$, and
$\sum_iw_i\bb c_i\bb c_i^\top=c_s^2I$ give
\begin{equation}\label{eq:LKSgenericEquilibrium}
 \sum_i a_i[\overline{\bb u}^{\,k}](\bb x_j)
 =1+2A_{\mathrm{LKS}}\Delta t\,c_s^2
                 \nabla_h\cdot\overline{\bb u}^{\,k}(\bb x_j).
\end{equation}
For $A_{\mathrm{LKS}}\ne0$, normalization therefore requires an exactly
divergence-free discrete velocity, which the recurrence does not enforce.
Our normalization-one perturbation bounds therefore remain conditional on this discrete divergence-free property. Uniform positivity is guaranteed whenever the prescribed velocity and its symmetric discrete gradient satisfy the sufficient bound
\[
 \begin{aligned}
 &\frac{|\bb c_i\cdot\overline{\bb u}^{\,k}(\bb x_j)|}{c_s^2}
 +\frac{|\overline{\bb u}^{\,k}(\bb x_j)|^2}{2c_s^2}\\
 &\quad{}+|A_{\mathrm{LKS}}|\Delta t\,
 \left|\bb c_i^\top\bigl(\nabla_h\overline{\bb u}^{\,k}
       +(\nabla_h\overline{\bb u}^{\,k})^\top\bigr)(\bb x_j)
       \bb c_i\right|\leq1-\delta,
 \qquad\text{for all }k,i,j,
 \end{aligned}
\]
with a fixed margin $\delta>0$. Because the directional velocity-square term omitted from the equilibrium bracket is nonnegative, this inequality ensures $a_i[\overline{\bb u}^{\,k}]\geq\delta w_i$, whereas small velocity magnitude alone is insufficient whenever the symmetric gradient correction is large. This inequality serves solely as a sufficient admissibility condition for the quantum encoding rather than a stability claim across every parameter regime of the LKS.

Whereas multi-step coherent propagation along an admissible prescribed trajectory obeys the query bound of Proposition~\ref{prop:LKSextension} under consecutive-step weights, our hybrid single-step circuit uses the current computed weight as its input and output scaling because the next-step equilibrium weights remain unknown prior to classical core readout.

For this prescribed trajectory, define $\bb g^k=(\mathcal A^k)^{-1/2}\bb f^k$ and
$B^k=(\mathcal A^{k+1})^{-1/2}M^k(\mathcal A^k)^{1/2}$.
Its two-time-level streaming ratio is
$\rho^k=\max_{i,j}\sqrt{a_i[\overline{\bb u}^{\,k}](\bb x_j)/a_i[\overline{\bb u}^{\,k+1}](\bb x_{\pi_i(j)})}$.
This ratio uses consecutive reference weights, whereas $\overline\rho^k$
in the hybrid single-step construction uses the current computed weight on both sides.

If $\sum_i a_i[\overline{\bb u}^{\,k}]\ne1$, the projection argument cannot be used unchanged:
at a node, the scaled full-relaxation collision is
$\sqrt{\bb a}\sqrt{\bb a}^{\,\top}$, where
$\bb a=(a_i[\overline{\bb u}^{\,k}](\bb x_j))_{i=0}^{q-1}$.
Its norm is $\sum_i a_i[\overline{\bb u}^{\,k}](\bb x_j)$
when the coefficients are positive. That factor must be included in the
normalization bound: with $s^k_j=\sum_i a_i[\overline{\bb u}^{\,k}](\bb x_j)$, one has
\[
 \|B^k\|\leq\rho^k\max_j s^k_j.
\]
Furthermore, recovery uses
$\|(\mathcal A^k)^{1/2}\|=\sqrt{\max_{i,j}a_i[\overline{\bb u}^{\,k}](\bb x_j)}$,
which need not be at most one in this case. These factors must be
retained when extending the normalized-coefficient estimates to a
trajectory with nonzero discrete divergence. In \eqref{eq:LKSequilibriumExplicit}, the term proportional to $A_{\mathrm{LKS}}\Delta t$ supplies the viscous stress correction that sets the nominal viscosity in \eqref{eq:LKSviscosity}; dividing $a_i[\overline{\bb u}^{\,k}](\bb x_j)$ nodewise by $s_j^k$ would restore an orthogonal projection at $\tau^*=1$, but would alter both the local mass update and the second-moment momentum flux whenever $\nabla_h\cdot\overline{\bb u}^{\,k}(\bb x_j)\ne0$.

\subsection{Current-step scaling and changes of weighted norm}
\label{subsec:TTweightChanges}

Because the hybrid circuit scales step $k$ solely by the current weight whereas multi-step error bounds telescope across consecutive time levels, the analysis distinguishes the current-weight operator $\overline B^k=(\mathcal A^k)^{-1/2}M^k(\mathcal A^k)^{1/2}$ from the two-time-level operator $B^k=(\mathcal A^{k+1})^{-1/2}M^k(\mathcal A^k)^{1/2}$. The exact reference weights used in these estimates need not be available to the algorithm. The underlying prescribed-coefficient norm bounds and block-encoding estimates are collected in Appendix~\ref{app:linearBE}.

Along an exact positive-coefficient trajectory, we set $\mathcal A^k=\diag(\bb a[\bb u^k])$ and $\overline B^k=(\mathcal A^k)^{-1/2}M^k(\mathcal A^k)^{1/2}$, so the prescribed-coefficient two-time-level operator in~\eqref{gtime} satisfies
\[
 B^k=(\mathcal A^{k+1})^{-1/2}(\mathcal A^k)^{1/2}
                    \overline B^k.
\]
Converting $\overline{\bb g}^{\,k+1}=\overline B^k\bb g^k$ via multiplication by the first diagonal ratio factor therefore yields $\bb g^{k+1}$. Although the hybrid loop evaluates the next-step weights only after physical core reconstruction, multi-step error growth still depends on the space-time coefficient ratios along the trajectory and their sensitivity to feedback perturbations.

For the ideal single-step readout in
\eqref{eq:TTscaledPhysicalOutput}, the success probability is
\[
 \mathbb P_{\mathrm{step}}^k
 =\frac{\|\bb y^k\|^2}
 { (\alpha_{\mathrm{hyb}}^k)^2\|\bb g_{\mathrm{TT}}^k\|^2}.
\]
The square root of this denominator is the normalization factor for each measured overlap.
Thus the cost depends on recovery and the scaled input norm as well
as the collision--streaming normalization. When positivity bounds are
uniform, the recovery-to-input conversion satisfies
\[
 \sqrt{a_{\max}}\,
 \frac{\|\bb g_{\mathrm{TT}}^k\|}{\|\bb f_{\mathrm{TT}}^k\|}
 \leq\sqrt{\frac{a_{\max}}{a_{\min}}}.
\]
Recovery and input scaling therefore remain part of the readout cost,
even for a normalization-one collision. Their implementation errors, together with overlap-sampling and core-fitting errors, enter the physical update defect in our nonlinear perturbation budget.

\subsection{Conditional error propagation for nonlinear TT feedback}\label{subsec:TTfeedbackError}

We now bound how local coefficient, operator, and core-fitting defects accumulate in the physical-scale feedback loop of Section~\ref{sec:TThybrid}, where equilibrium weights are recomputed from the compressed state at every step. Evaluating the exact coefficients $\bb a[\bb u^k]$ along a reference trajectory $\bb f^k$, we assume that these reference coefficients satisfy the positivity and unit-sum conditions in~\eqref{eq:LKScoefficientConditions} and choose $\Lambda\geq\max\{1,\max_k\|B^k\|\}$ for the corresponding two-time-level scaled operators $B^k$. We measure the state error in the reference energy norm,
\[
 e_k=\|(\mathcal A^k)^{-1/2}
                  (\bb f_{\mathrm{TT}}^{k}-\bb f^k)\|,
\]
which serves solely as an analytical weight and need not be known to the algorithm; the resulting estimate is a conditional perturbation bound rather than a guarantee that a prescribed TT rank cap and ALS sweep budget attain the required local defect.

To separate feedback sensitivity from single-step update errors, suppose on the states under consideration
\begin{equation}\label{eq:LKSfeedbackLipschitz}
 \|\bb a[\bb u(\bb v)]-\bb a[\bb u(\bb w)]\|\leq L_a\|\bb v-\bb w\|,
 \qquad \|E_I\bb f_{\mathrm{TT}}^{k}\|_\infty\leq\varrho_{\max}.
\end{equation}
Here the Euclidean coefficient norm spans all discrete directions and spatial nodes, while the Lipschitz constant and uniform nodal density bound hold across all considered states. We let $d_k^f$ bound the total defect of the physical update executed with the computed coefficients $\bb a_{\mathrm{TT}}^{k}$,
\[
 \bb f_{\mathrm{TT}}^{k+1}
 =P\diag(\bb a_{\mathrm{TT}}^{k})E_I^\top E_I\bb f_{\mathrm{TT}}^{k}+\bb d^k,
 \qquad \|\bb d^k\|\leq d_k^f.
\]
This defect includes moment-contraction errors in applying the physical update with frozen coefficients, together with errors in multiplication, streaming, and final rounding. For the hybrid implementation, it also includes scaled-state preparation, quantum evolution and recovery errors, and the accepted core-fitting residual with its sampling uncertainty. The coefficient evaluation budget separately accounts for errors in the moment calculations used to form the coefficients.

\begin{proposition}[Error bound with nonlinear TT feedback]\label{prop:TTnonlinear}
 Under these assumptions and the coefficient error budget of
 Section~\ref{subsec:TTcoefficients}, set
\[
 \chi=\Lambda+\frac{\varrho_{\max}L_a}{\sqrt{a_*}},\qquad
 b_k=\frac{\varrho_{\max}\epsilon_a^k+d_k^f}{\sqrt{a_*}}.
\]
 Then
\begin{equation}\label{eq:TTnonlinearError}
 e_{k+1}\leq\chi e_k+b_k,\qquad
 \|\bb f_{\mathrm{TT}}^{N_t}-\bb f^{N_t}\|
 \leq e_{N_t}\leq\chi^{N_t}e_0+
                \sum_{k=0}^{N_t-1}\chi^{N_t-1-k}b_k.
\end{equation}
 If the right-hand side is less than $\|\bb f^{N_t}\|$, twice its
 ratio to $\|\bb f^{N_t}\|$ bounds the normalized distribution error.
\end{proposition}
\begin{proof}
 Subtracting the reference update separates linear propagation from coefficient-feedback and local update defects according to
\begin{align*}
 \bb f_{\mathrm{TT}}^{k+1}-\bb f^{k+1}
 ={}&M^k(\bb f_{\mathrm{TT}}^{k}-\bb f^k)\\
 &+P\diag(\bb a_{\mathrm{TT}}^{k}-\bb a[\bb u^k])E_I^\top E_I\bb f_{\mathrm{TT}}^{k}
 +\bb d^k.
\end{align*}
Applying the reference energy scaling at time level $k+1$ bounds the first term by $\Lambda e_k$ and the second term by $a_*^{-1/2}\varrho_{\max}\|\bb a_{\mathrm{TT}}^{k}-\bb a[\bb u^k]\|$, where the coefficient-evaluation tolerance, the Lipschitz condition~\eqref{eq:LKSfeedbackLipschitz}, and the unit upper bound on the normalized reference weights yield
\[
 \|\bb a_{\mathrm{TT}}^{k}-\bb a[\bb u^k]\|
 \leq\epsilon_a^k+L_a\|\bb f_{\mathrm{TT}}^{k}-\bb f^k\|
 \leq\epsilon_a^k+L_a e_k.
\]
Bounding the scaled local defect by $d_k^f/\sqrt{a_*}$ and iterating across time steps establishes~\eqref{eq:TTnonlinearError}, while physical-scale norm recovery and normalization follow directly from Proposition~\ref{prop:TTmarching} in Appendix~\ref{app:linearBE}.
\end{proof}

The feedback Lipschitz constant can be expressed in terms of identifiable flow and lattice quantities.
If both densities are at least $\varrho_{\min}$ and pointwise velocities
have magnitude at most $U$, then
\[
 \|\bb u(\bb v)-\bb u(\bb w)\|
 \leq\frac{\|\bb C\|+U\sqrt q}{\varrho_{\min}}\|\bb v-\bb w\|.
\]
If the equilibrium map is Lipschitz with constant $L_{\mathrm{eq}}$
in the sum of the velocity and gradient norms, one admissible choice is
\[
 L_a=L_{\mathrm{eq}}(1+\|\nabla_h\|)
             \frac{\|\bb C\|+U\sqrt q}{\varrho_{\min}}.
\]
In particular, if $\epsilon_{\mathrm{eval}}^k$ bounds the additional
velocity and length-weighted gradient errors from TT evaluation, then
\[
 \|\bb u_{\mathrm{TT}}^{k}-\bb u^k\|
 +\Delta x\|\nabla_h\bb u_{\mathrm{TT}}^{k}-\nabla_h\bb u^k\|
 \leq(1+\Delta x\|\nabla_h\|)
 \frac{\|\bb C\|+U\sqrt q}{\varrho_{\min}}e_k
 +\epsilon_{\mathrm{eval}}^k.
\]
Together with~\eqref{eq:TTnonlinearError}, this macroscopic sensitivity estimate bounds the output errors in~\eqref{eq:LKSoutputRequirements} and guides tolerance allocation.
In the weakly compressible low-Mach regime ($U\ll c_s$ with $\varrho^k\approx\varrho_{\mathrm{ref}}$), $\|\bb f^k\|$ is dominated by the $\mathcal O(\varrho_{\mathrm{ref}})$ zeroth-moment background whereas the momentum $\bb m^k=\bb C\bb f^k$ arises from $\mathcal O(\varrho_{\mathrm{ref}}U)$ signed cancellations across opposite lattice velocities. A small relative distribution defect can therefore be amplified by $\mathcal O(c/U)$ in relative velocity error, mirroring macro-micro moment sensitivity in classical low-rank kinetic approximations \cite{coughlin2024robust,dektor2026interpolatory}.
The estimate remains conditional: because the discrete gradient norm $\|\nabla_h\|$ grows as $(\Delta x)^{-1}$ under mesh refinement, \eqref{eq:TTnonlinearError} is a fixed-grid perturbation bound rather than a grid-independent nonlinear stability theorem, and mass and momentum must be monitored explicitly since generic TT rounding does not enforce discrete conservation laws.

\section{End-to-end resource requirements}\label{sec:resources}\label{subsec:TTtotalComplexity}

The error analysis specifies what must be accurate; the resource model now counts the work required to achieve that accuracy. Each reconstructed core requires repeated single-step circuits and overlap measurements, while coefficient evaluation, TT arithmetic, and residual checks add classical work. We count these operations separately before combining them. The resulting bounds depend on attained ranks, sweep counts, precision budgets, and accepted residuals; a small number of stored parameters alone does not fix the work per step.

\subsection{Time-dependent quantum execution costs}\label{subsec:quantumExecCost}

Because Hadamard overlap estimation restarts from the stored classical cores at step $k$ without replaying earlier steps, total quantum circuit executions accumulate across local core entries and alternating sweeps within each physical time step. To account for this work, let $S_k$ be the number of fitting
sweeps at time $k$ and let $d_s^{k,\ell}$ be the number of entries in core
$s$ during sweep $\ell$. Let $N_{\mathrm{meas}}^{k,\ell,s,t}$ denote
the number of Hadamard-test runs used for its $t$-th overlap. In gate
counts, the measurement contribution is
\begin{equation}\label{eq:TThybridMeasurementCost}
 C_{\mathrm{meas}}
 =\sum_{k=0}^{N_t-1}\sum_{\ell=1}^{S_k}\sum_{s=1}^{p}
   \sum_{t=1}^{d_s^{k,\ell}}N_{\mathrm{meas}}^{k,\ell,s,t}
 \left(C_{\mathrm{in}}^k+C_{\mathrm{step}}^k
                         +C_{\mathrm{test}}^{k,\ell,s,t}\right).
\end{equation}
Here $C_{\mathrm{in}}^k$ counts the gates for controlled input preparation,
$C_{\mathrm{step}}^k$ counts the gates for weighted evolution and recovery, and
$C_{\mathrm{test}}^{k,\ell,s,t}$ counts the gates for inverse test-state preparation
together with the remaining Hadamard gates. The input and step gate counts depend only on the physical time step because the equilibrium coefficients can vary between steps while remaining fixed throughout the fitting sweeps of a given step. By contrast, candidate ranks and overlap-estimation precision can vary across fitting sweeps, and the overlap normalization
$\alpha_{\mathrm{hyb}}^k n_{g,k}$ directly enters the required shot count.

\begin{remark}[Classical TT storage avoids replaying earlier time steps]
\label{rem:TTnoHistoryReplay}
The accepted classical cores store $\bb f_{\mathrm{TT}}^k$ with its
physical scale. Each Hadamard test prepares its input from these cores
and executes only the current step, recovery, and test-state circuit.
Earlier quantum time steps are not replayed.

Thus the single-run cost in \eqref{eq:TThybridMeasurementCost} is
\[
 C_{\mathrm{in}}^k+C_{\mathrm{step}}^k
                  +C_{\mathrm{test}}^{k,\ell,s,t},
\]
without summing earlier evolution costs. Repeated shots and TT loading
are still required; their costs may vary with ranks and precision.
This step-local restart from stored cores differs from preparing each input by a coherent prefix starting
at the initial time, but does not by itself imply a lower total cost.

\end{remark}

\subsection{Sampling cost of a local core update}\label{subsec:localSamplingCost}

Converting the execution count in \eqref{eq:TThybridMeasurementCost} into an explicit shot budget requires bounding the statistical repetitions $N_{\mathrm{meas}}^{k,\ell,s,t}$ of the elementary Hadamard-test estimator for one local core update. Define the known amplitude scale
\[
 \mathcal N_k=\alpha_{\mathrm{hyb}}^k n_{g,k}.
\]
For one update of core $s$ in sweep $\ell$, we abbreviate
$d=d_s^{k,\ell}$ and allocate an absolute Euclidean sampling-error budget
$\eta_{k,\ell,s}>0$ to the vectorized core together with a failure probability
$\delta_{k,\ell,s}\in(0,1)$. Because each Hadamard-test outcome for overlap $t$ takes
values $\pm1$ with expectation
$\mu_t=(\bb\theta_s)_t/\mathcal N_k$, we estimate each expectation to
absolute error
\[
 \epsilon_{\mu}=\min\left\{1,
             \frac{\eta_{k,\ell,s}}{\mathcal N_k\sqrt d}\right\}.
\]
Hoeffding's inequality and a union bound over the $d$ entries show that
it suffices to take, for each entry,
\begin{equation}\label{eq:TTshotsPerEntry}
 N_{\mathrm{meas}}^{k,\ell,s,t}
 =\left\lceil\frac{2}{\epsilon_\mu^2}
                  \log\frac{2d}{\delta_{k,\ell,s}}\right\rceil.
\end{equation}
Then, with probability at least $1-\delta_{k,\ell,s}$, the reconstructed
core has sampling error at most $\eta_{k,\ell,s}$, since
\[
 \|\widehat{\bb\theta}_s-\bb\theta_s\|
 \leq\mathcal N_k\sqrt d\,\epsilon_\mu
 \leq\eta_{k,\ell,s}.
\]
When the remaining candidate cores are held fixed, the orthonormality of
$\mathsf X_s$ transfers the same Euclidean error bound to the updated TT distribution.
Across adaptive sweeps, preallocating per-update failure budgets with total at most $\delta_{\mathrm{fail}}$ allows conditional concentration and a union bound to control all core-sampling errors, while complete step certification also allocates residual-test failure budgets together with separate circuit and scaling error budgets.

Let $C_{\mathrm{run}}^{k,\ell,s}$ be the largest full single-run gate
count among the $d_s^{k,\ell}$ overlap circuits for this core. Substituting
\eqref{eq:TTshotsPerEntry} into \eqref{eq:TThybridMeasurementCost} yields
\begin{equation}\label{eq:TThybridSamplingBound}
\begin{aligned}
 C_{\mathrm{meas}}=\mathcal O\!\Bigg(
 \sum_{k=0}^{N_t-1}\sum_{\ell=1}^{S_k}\sum_{s=1}^{p}
 &C_{\mathrm{run}}^{k,\ell,s}d_s^{k,\ell}
 \max\!\left\{1,
 \frac{\mathcal N_k^2 d_s^{k,\ell}}{\eta_{k,\ell,s}^2}\right\}\\
 &\hspace{1em}\times
 \log\frac{2d_s^{k,\ell}}{\delta_{k,\ell,s}}\Bigg).
\end{aligned}
\end{equation}
For independent Hadamard tests in the small-error regime, a whole-core
update costs $\mathcal O(d^2\mathcal N_k^2/\eta_{k,\ell,s}^2)$ samples
up to confidence logarithms: one factor of $d$ counts the core entries, and the second comes from splitting the Euclidean core tolerance $\eta_{k,\ell,s}$ across $d$ entries. Bounded TT ranks therefore replace $\mathcal O(qN)$ grid amplitudes by $\sum_{s=1}^p d_s^{k,\ell}\leq pn_{\max}r^2$ fitted parameters, removing explicit linear dependence on $N$ from the number of fitted parameters when $p=\mathcal O(\log N)$. This parameter count does not remove grid-size dependence from the sampling budget: $\mathcal N_k/\eta_{k,\ell,s}$ may still grow with $N$ under a fixed absolute core tolerance. Inverse-square overlap sampling and the composite scale $\mathcal N_k=\alpha_{\mathrm{hyb}}^k n_{g,k}$ (combining input energy scaling, discrete-divergence normalization, and amplitude recovery) remain in the measurement cost. The relative core tolerances used in Section~\ref{subsec:TTnumericalExample} scale with the distribution norm; fixed absolute tolerances describe a different resource requirement.

\subsection{Classical work, storage, and total complexity}\label{subsec:totalComplexity}

Completing the end-to-end resource model requires combining the quantum sampling executions of Sections~\ref{subsec:quantumExecCost}--\ref{subsec:localSamplingCost} with classical TT arithmetic, residual verification, and memory requirements without double counting. We first count quantum gates and classical arithmetic operations separately. The total quantum gate count is
\begin{equation}\label{eq:TThybridTotalQuantumCost}
 C_{\mathrm{quantum}}=C_{\mathrm{meas}}+C_{\mathrm{check}},
\end{equation}
where $C_{\mathrm{check}}$ counts the quantum gates of all additional executions for residual or other acceptance tests; when these tests are performed entirely by classical TT contractions, their cost belongs to the classical work instead. Let $W_{\mathrm{init}}$ denote initial tensorization and TT construction work, where full-array initialization reads $qN$ values whereas structured initial data may admit direct core construction. With each per-step term summing every invocation at time level $k$, the classical operation count is
\begin{equation}\label{eq:TThybridTotalClassicalCost}
 W_{\mathrm{classical}}
 =W_{\mathrm{init}}+\sum_{k=0}^{N_t-1}
 \left(W_{\mathrm{coeff}}^k+W_{\mathrm{scale}}^k
       +W_{\mathrm{compile}}^k+W_{\mathrm{fit}}^k
       +W_{\mathrm{check}}^k\right).
\end{equation}
Here the five per-step terms account respectively for moments, density division, gradients, and equilibrium evaluation; forming inverse-square-root fields and scaled TT inputs; state, oracle, and test-vector circuit compilation; sample processing, core updates, orthogonalization, and rounding; and classical residual and admissibility tests.

In storage, a TT distribution of maximum mode size $n_{\max}$ and rank $r$ requires $\mathcal O(pn_{\max}r^2)$ numbers, with at most $n_{\max}r^2$ entries per core. Peak working memory additionally holds coefficient and scaled-input cores, intermediate products, contraction environments, and workspaces. Because local sampling tolerances alone do not bound the accumulated physical error, the propagation analysis in Section~\ref{subsec:TTfeedbackError} combines coefficient, scaling, circuit, sampling, and fitting defects, leaving all resource bounds conditional on attained ranks and accepted residuals.

To state a single combined work bound, we adopt a unit-cost model for each classical arithmetic operation and elementary quantum gate at the specified precision. We then let $C_{\mathrm{I/O}}$ count additional elementary measurement, reset, and data-transfer operations not already included in \eqref{eq:TThybridTotalQuantumCost}--\eqref{eq:TThybridTotalClassicalCost}. Counting each operation once gives the total hybrid work
\begin{equation}\label{eq:TThybridFinalTotalCost}
 C_{\text{total}}
 =W_{\mathrm{classical}}+C_{\mathrm{meas}}
                         +C_{\mathrm{check}}+C_{\mathrm{I/O}}.
\end{equation}
Elapsed-time estimates instead weight classical arithmetic, quantum gates, measurements, and communication by their hardware implementation times.

If uniform bounds $S_k\leq S$, $d_s^{k,\ell}\leq d_{\max}$, $C_{\mathrm{run}}^{k,\ell,s}\leq C_{\max}$, $\mathcal N_k/\eta_{k,\ell,s}\leq H$, and $\delta_{k,\ell,s}\geq\delta_{\min}>0$ hold across the performed updates, substituting \eqref{eq:TThybridSamplingBound} into \eqref{eq:TThybridFinalTotalCost} yields the conditional total complexity bound
\begin{equation}\label{eq:TThybridFinalTotalBound}
\begin{aligned}
 C_{\text{total}}=\mathcal O\!\Bigg(&W_{\mathrm{classical}}
                 +C_{\mathrm{check}}+C_{\mathrm{I/O}}\\
 &+N_t S p C_{\max}d_{\max}\max\{1,H^2d_{\max}\}
          \log\frac{2d_{\max}}{\delta_{\min}}\Bigg).
\end{aligned}
\end{equation}
Here $S$, $d_{\max}$, $C_{\max}$, and $H$ may still depend on grid size, time horizon, and target accuracy; when parameters vary across steps or cores, \eqref{eq:TThybridFinalTotalCost} is used directly with \eqref{eq:TThybridSamplingBound}. Any comparison with purely classical TT propagation requires matched physical accuracy and confidence.

\section{Numerical evidence for discretization, compression, feedback, and readout}\label{sec:numerics}\label{subsec:nonlinearExperiments}

The numerical tests examine whether the information retained by the hybrid interface is sufficient for accurate flow feedback and what precision its readout demands. We first check the underlying discretization and the algebraic change of coordinates on a periodic Taylor--Green vortex. A classical three-dimensional cavity calculation then tests wall-bounded flow and static quantized TT (QTT) compression. The final two tests introduce repeated TT reconstruction and simulated sampling noise, respectively, to separate compression error from measurement error.

All four studies use classical computations or statistical simulations. The feedback tests use dense reference operations where specified and hold the ranks and number of fitting sweeps fixed; they report achieved distribution and velocity errors. They do not certify all acceptance conditions of the adaptive algorithm, including its joint velocity--gradient tolerance. Single-step coefficient checks also leave the reference-trajectory assumptions of Section~\ref{sec:hybridAnalysis} in place. These diagnostics establish the behavior of individual parts of the interface, with full circuit execution and a matched-cost comparison remaining implementation tasks.

\subsection{Two-dimensional Taylor--Green vortex verification}
\label{subsec:TaylorGreenBenchmark}

Before introducing low-rank truncation or measurement noise, we verify the baseline full-relaxation LKS discretization and confirm that the single-step current-weight transformation of Section~\ref{subsec:TTquantumStep} is algebraically consistent with direct physical updates on the periodic two-dimensional Taylor--Green benchmark of Xiao et al.~\cite{xiao2025latticekinetic}.

On $[-L,L)^2$, let $\kappa=\pi/L$. The reference velocity and the
associated weakly compressible density initialization are
\begin{equation}\label{eq:TaylorGreenReference}
\begin{aligned}
 u_x(x,y,t)&=-u_0\cos(\kappa x)\sin(\kappa y)
                         \exp(-2\nu\kappa^2t),\\
 u_y(x,y,t)&= u_0\sin(\kappa x)\cos(\kappa y)
                         \exp(-2\nu\kappa^2t),\\
 \varrho(x,y,t)&=\varrho_0-
 \frac{\varrho_0u_0^2}{4c_s^2}
 [\cos(2\kappa x)+\cos(2\kappa y)]\exp(-4\nu\kappa^2t).
\end{aligned}
\end{equation}
Here the velocity field solves the continuum incompressible Navier--Stokes equations while the density
expression supplies a pressure-compatible kinetic initialization;
we do not regard these analytical fields as an exact solution of the discrete LKS.
We use D2Q9 with $\Delta x=\Delta t=1$, $c_s^2=1/3$,
$\varrho_0=1$, and $\nu=0.08$, giving $A_{\mathrm{LKS}}=0.39$ in
\eqref{eq:LKSviscosity}. For $n=8,16,32,64$ nodes per coordinate,
we set $L=n/2$ and $u_0=1.6/n$ to fix $Re=u_0L/\nu=10$,
for which the common dimensionless final time $t^*=u_0T/L=1$ corresponds to step
counts $N_t=20,80,320,1280$, respectively. We initialize
$f_i^0=\varrho^0a_i[\bb u^0]$ from the gradient-corrected equilibrium
\eqref{eq:LKSequilibriumExplicit}, with spatial gradients evaluated by the
periodic centered differences in \eqref{eq:LKSdiscreteGradient}.

\begin{figure}[htbp]
\centering
\subfigure[Velocity error under refinement]{
 \includegraphics[width=0.47\linewidth]{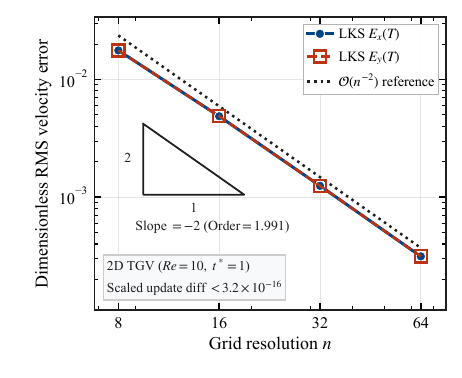}
}
\hfill
\subfigure[Contours of $u_x/u_0$ on the $64\times64$ grid]{
 \includegraphics[width=0.47\linewidth]{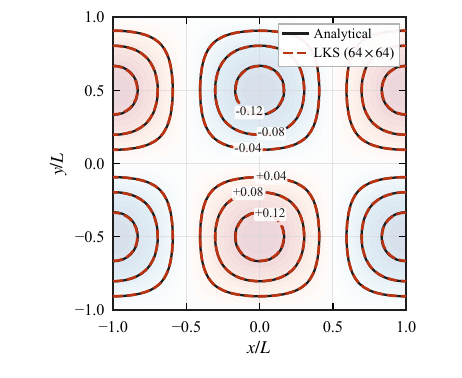}
}
\caption{Periodic Taylor--Green vortex at $t^*=1$. In (b), the contour levels are $\pm0.04,\pm0.08,\pm0.12$ over the shaded analytical field; solid and dashed curves denote the analytical velocity and the computed LKS solution, respectively.}
\label{fig:TaylorGreenRefinement}
\end{figure}

For each velocity component $\gamma\in\{x,y\}$, we report the
dimensionless root-mean-square error
\[
 E_\gamma(T)=\frac{1}{u_0}
 \left(\frac{1}{n^2}\sum_j
 |u_{\gamma,\mathrm{LKS}}(\bb x_j,T)
       -u_\gamma(\bb x_j,T)|^2\right)^{1/2}.
\]
This normalization differs from division by the norm of the decaying
reference velocity. Under this fixed-$Re$, decreasing-Mach-number refinement, the discrete velocity field converges toward the analytical solution at nearly second order, with $\log_2(E_x(n)/E_x(2n))$ reaching $1.991$ on the $64\times64$ grid as the normalized error drops to $3.1399\times10^{-4}$ (Table~\ref{tab:TaylorGreenRefinement} and Figure~\ref{fig:TaylorGreenRefinement}).
At every step for $n\leq32$ and at the initial, final, and every eighth step for $n=64$, a separate evaluation of the current-weight collision, streaming, and recovery reproduces the direct physical update to relative error below $3.2\times10^{-16}$.
Across the trajectory, all evaluated equilibrium coefficients and densities remain positive ($a_i\geq0.0144$ and $\varrho\geq0.94$), and the relative mass drift decreases from $6.33\times10^{-5}$ to
$3.81\times10^{-8}$ over the four grids even though exact mass conservation is
not imposed by the discrete equilibrium coefficient sum in \eqref{eq:LKSgenericEquilibrium}. These periodic baseline results fix the discretization error floor before we examine three-dimensional wall-bounded flow and static QTT compressibility in Section~\ref{subsec:cavity3d}.

\begin{table}[htbp]
\centering\small
\caption{Independently computed two-dimensional Taylor--Green vortex
benchmark at $Re=10$ and $t^*=1$. The observed order refers to $E_x$;
$E_y$ agrees with $E_x$ to the displayed precision.}
\label{tab:TaylorGreenRefinement}
\begin{tabular}{rrrrrr}
\toprule
$n$ & $u_0$ & $N_t$ & $E_x$ & $E_y$ & Order\\
\midrule
8  & 0.200 & 20   & $1.7675\times10^{-2}$ & $1.7675\times10^{-2}$ & ---\\
16 & 0.100 & 80   & $4.8744\times10^{-3}$ & $4.8744\times10^{-3}$ & 1.858\\
32 & 0.050 & 320  & $1.2485\times10^{-3}$ & $1.2485\times10^{-3}$ & 1.965\\
64 & 0.025 & 1280 & $3.1399\times10^{-4}$ & $3.1399\times10^{-4}$ & 1.991\\
\bottomrule
\end{tabular}
\end{table}

\FloatBarrier
\subsection{Classical three-dimensional cavity flow and static QTT snapshot compression}
\label{subsec:cavity3d}

The cavity calculation tests two properties that can be examined independently of a feedback trajectory: the classical LKS behavior with walls and the compressibility of a three-dimensional distribution snapshot. We use the D3Q27 lid-driven cavity at $Re=100$ following Xiao et al.~\cite{xiao2025latticekinetic}, together with static QTT compression of one computed snapshot. The top lid $z=L$ translates with velocity $(U,0,0)$ while the other five walls remain stationary. In lattice units, we take $\Delta x=\Delta t=1$, $U=0.1$, and $L=n$ on an $n^3$ cell-centered grid with $n=16,24,32,48$, setting $\nu=UL/100$ and $A_{\mathrm{LKS}}=3/4-9\nu/2$ from \eqref{eq:LKSviscosity}. All runs start from the rest-state weights $f_i^0=w_i$.

Unlike the periodic tests in Sections~\ref{subsec:TaylorGreenBenchmark}, \ref{subsec:TTfeedbackRanks}, and~\ref{subsec:TTnumericalExample}, this experiment requires a wall
closure. We use halfway bounce-back with a moving-wall correction
\cite{kruger2017lattice}. For a pull-streaming link whose source lies
outside the cavity, we set
\begin{equation}\label{eq:cavityWallRule}
 f_i^{k+1}(\bb x)
 =f_{\bar i}^{\mathrm{col},k}(\bb x)
  +\frac{2w_i\varrho^k(\bb x)}{c_s^2}\bb c_i\cdot\bb u_w,
 \qquad \bb c_{\bar i}=-\bb c_i,
\end{equation}
where $\bb u_w=(U,0,0)$ at the lid and zero elsewhere.
Corner links crossing both the lid and a stationary wall use the stationary-wall
velocity. Interior velocity gradients use centered differences, whereas boundary-adjacent centered differences at a boundary face use the ghost
velocity $2\bb u_w-\bb u_{\mathrm{adj}}$ formed from the adjacent fluid velocity $\bb u_{\mathrm{adj}}$. We retain the original
gradient-corrected equilibrium coefficients without normalizing their sum or
correcting total mass, forming a self-contained boundary treatment
independent of the wall closure in the cited benchmark.

We stop when the velocity residual
\[
 R_{100}(k)=\frac{\|\bb u^k-\bb u^{k-100}\|}{\|\bb u^k\|}
 <10^{-7}
\]
holds at three consecutive checks, spaced 100 steps apart after the
first 1000 steps. This tests velocity stationarity, not stationarity
of the density or distribution. As a separate discretization comparison,
we also run standard quadratic-equilibrium D3Q27 BGK-LBM on $32^3$
cells with the same walls, $U$, and nominal $\nu$, using
$\tau^*=1/2+3\nu=0.596$.

\begin{figure}[htbp]
\centering
\subfigure[$x$--$z$ plane, $y/L=1/2$]{
 \includegraphics[width=0.31\linewidth]{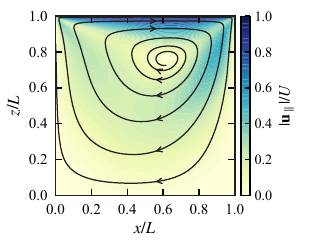}
}
\hfill
\subfigure[$y$--$z$ plane, $x/L=1/2$]{
 \includegraphics[width=0.31\linewidth]{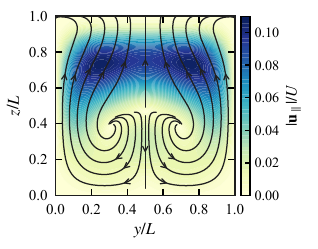}
}
\hfill
\subfigure[$x$--$y$ plane, $z/L=1/2$]{
 \includegraphics[width=0.31\linewidth]{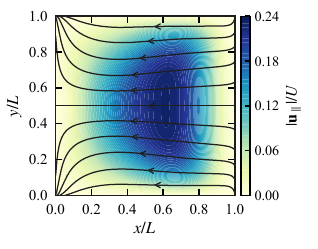}
}
\par\medskip
\subfigure[Vertical centerline velocity]{
 \includegraphics[width=0.47\linewidth]{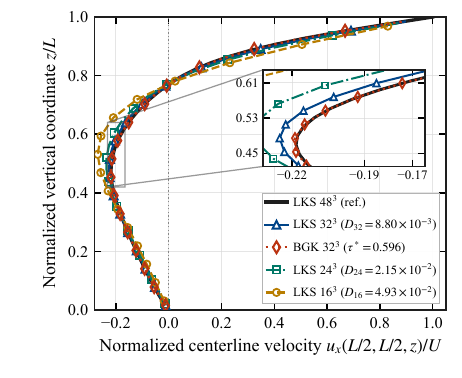}
}
\hfill
\subfigure[Velocity convergence]{
 \includegraphics[width=0.47\linewidth]{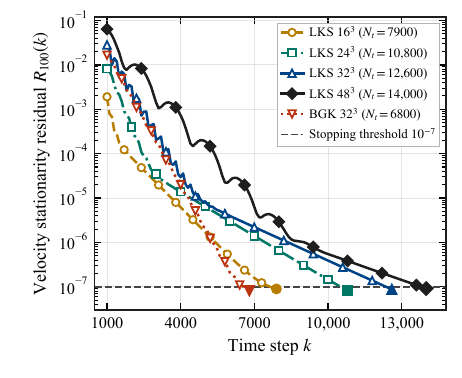}
}
\caption{Classical three-dimensional cavity computations at $Re=100$.
The upper panels show streamlines of the in-plane velocity projections
on the $48^3$ grid; they need not be three-dimensional particle paths.
The lower panels compare mesh refinement with an independent BGK-LBM
run and show the velocity stopping criterion.}
\label{fig:cavity3d}
\end{figure}

At steady velocity residual, the full-relaxation D3Q27 solver develops the primary recirculation in the central $x$--$z$ plane together with symmetric transverse secondary structures, as seen in the $48^3$ in-plane velocity projections, centerline profiles, and convergence histories of Figure~\ref{fig:cavity3d}. Across the $16^3$, $24^3$, and $32^3$ grids, the RMS centerline difference $D_n$ relative to the $48^3$ LKS curve decreases monotonically from $4.93\times10^{-2}$ to $8.80\times10^{-3}$ (Table~\ref{tab:cavity3d}), where velocities are normalized by $U$, averaged over the four adjacent cell-center values, and linearly interpolated to 181 points in $0.05\leq z/L\leq0.95$ at $x=y=L/2$. This metric is a refinement diagnostic rather than an error against an exact solution; in the same metric, the $32^3$ BGK comparison differs from the finest LKS profile by $1.52\times10^{-3}$.

\begin{table}[htbp]
\centering\small
\caption{Three-dimensional cavity at $Re=100$. Here $N_t$ is the stopping
step, $D_n$ is the centerline difference from the $48^3$ LKS result,
and $E_{\mathrm{mass}}(20)$ is relative mass drift at the common time
$t^*=kU/L=20$.}
\label{tab:cavity3d}
\begin{tabular}{rrrrr}
\toprule
Grid & $A_{\mathrm{LKS}}$ & $N_t$ & $D_n$ & $E_{\mathrm{mass}}(20)$\\
\midrule
$16^3$ & 0.678 & 7900 & $4.93\times10^{-2}$ & $1.47\times10^{-2}$ \\
$24^3$ & 0.642 & 10800 & $2.15\times10^{-2}$ & $1.14\times10^{-2}$ \\
$32^3$ & 0.606 & 12600 & $8.80\times10^{-3}$ & $9.67\times10^{-3}$ \\
$48^3$ & 0.534 & 14000 & --- & $7.90\times10^{-3}$ \\
\bottomrule
\end{tabular}
\end{table}

The tabulated relative mass drift is defined by $|\sum_j\varrho_j^k/n^3-1|$.
Although centerline differences decrease under refinement, mass drift
remains visible because neither the wall treatment nor the gradient-corrected coefficient
sum imposes exact mass conservation. Positivity checks at scheduled
diagnostic checkpoints give $a_i\geq0.00321$ and
$\varrho\geq0.937$ across all LKS runs.
On identical current inputs, direct physical updates and weighted collision
followed by the same wall rule and physical recovery agree to
relative error below $3.3\times10^{-16}$. These positivity and weighted-update checks are evaluated at the first step,
every 1000 steps, and at termination. They verify the single-step
algebra rather than a quantum circuit implementation.

For the final $32^3$ distribution, TT-SVD with modes $(27,2,\ldots,2)$ and 15 binary spatial modes stores 22253 core entries at rank cap 32, compared with 884736 distribution values. The relative distribution and velocity errors are $3.45\times10^{-4}$ and $6.88\times10^{-3}$, respectively. This single-snapshot result measures spatial compression at a specified velocity accuracy. It leaves open how repeated compression changes the coefficients and flow over time, which motivates the periodic feedback test in Section~\ref{subsec:TTfeedbackRanks}. Extending the quantum construction to this cavity would additionally require an operator for the moving-wall term in \eqref{eq:cavityWallRule}; the periodic normalization and query bounds do not extend to this wall closure without additional analysis.

\FloatBarrier
\subsection{Exact-overlap TT feedback and low-Mach rank sensitivity}\label{subsec:TTfeedbackRanks}

To test repeated nonlinear feedback, we return to the periodic Taylor--Green vortex of Section~\ref{subsec:TaylorGreenBenchmark} on a $16\times16$ grid and remove measurement noise by evaluating overlaps exactly. We tensorize the distribution with mode sizes $(9,2,2,2,2,2,2,2,2)$, placing the direction first, followed by the four binary digits of each spatial coordinate, and initialize by TT-SVD with rank caps $4,6,8,12$. At every step, the current compressed distribution supplies the nonlinear coefficients, and two alternating core sweeps fit the resulting physical target, one in each direction, as in Section~\ref{subsec:TTreadout}. Exact classical contractions supply the signed overlaps. Dense coefficient evaluation and target construction isolate the effects of compression and feedback; the test therefore measures trajectory accuracy rather than the cost of TT coefficient access or quantum execution.

Let $\bb f_{\mathrm{LKS}}^k$ and $\bb u_{\mathrm{LKS}}^k$ denote
the uncompressed discrete reference initialized from the uncompressed
data. We compare the compressed trajectory with this reference through the relative errors
\[
 E_{f,\mathrm{TT}}^k=
 \frac{\|\bb f_{\mathrm{TT}}^k-\bb f_{\mathrm{LKS}}^k\|}
      {\|\bb f_{\mathrm{LKS}}^k\|},\qquad
 E_{u,\mathrm{TT}}^k=
 \frac{\|\bb u_{\mathrm{TT}}^k-\bb u_{\mathrm{LKS}}^k\|}
      {\|\bb u_{\mathrm{LKS}}^k\|}.
\]
Increasing the rank cap from 8 to 12 improves agreement with the uncompressed discrete velocity while retaining a reduction in distribution storage (Table~\ref{tab:TaylorGreenTT}). Rank eight uses 796 entries in place of 2304 distribution values and reaches a final relative velocity error of $3.53\times10^{-3}$; rank twelve uses 1437 entries and reaches $1.03\times10^{-3}$. The table also reports $E_x(T)$ against the analytical velocity, which includes the underlying discretization error. The comparisons against uncompressed LKS include initial compression and the errors left by two fitting sweeps at each step; the fixed sweep count records achieved accuracy rather than convergence to a prescribed residual tolerance.

\begin{table}[htbp]
\centering\small
\caption{TT feedback for the Taylor--Green vortex on a $16\times16$
grid after 80 steps. Storage counts classical distribution-core entries.
All overlaps are evaluated exactly, without simulated shot noise.}
\label{tab:TaylorGreenTT}
\begin{tabular}{rrrrrr}
\toprule
Rank cap & Storage & $E_{f,\mathrm{TT}}^0$ & $E_{f,\mathrm{TT}}^{80}$
 & $E_{u,\mathrm{TT}}^{80}$ & $E_x(T)$\\
\midrule
4 & 248 & $2.28\times10^{-2}$ & $9.47\times10^{-3}$ & $1.000$ & $6.95\times10^{-2}$\\
6 & 482 & $5.15\times10^{-3}$ & $4.29\times10^{-4}$ & $1.90\times10^{-2}$ & $6.16\times10^{-3}$\\
8 & 796 & $1.25\times10^{-3}$ & $1.03\times10^{-4}$ & $3.53\times10^{-3}$ & $4.93\times10^{-3}$\\
12 & 1437 & $4.73\times10^{-6}$ & $1.77\times10^{-5}$ & $1.03\times10^{-3}$ & $4.87\times10^{-3}$\\
\bottomrule
\end{tabular}
\end{table}

\begin{figure}[htbp]
\centering
\subfigure[Velocity error vs.\ uncompressed trajectory]{
 \includegraphics[width=0.47\linewidth]{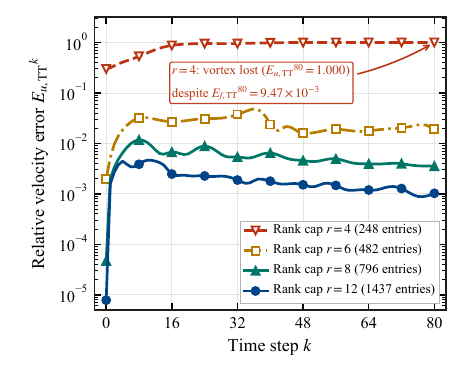}
}
\hfill
\subfigure[Final velocity profile at $x=0$]{
 \includegraphics[width=0.47\linewidth]{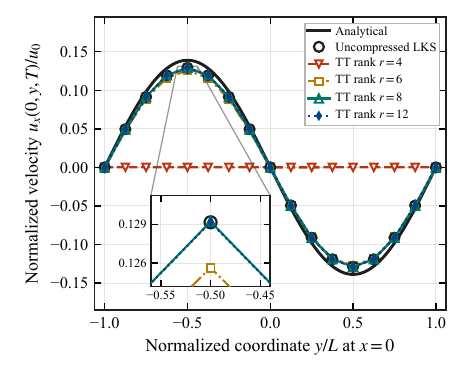}
}
\caption{Exact-overlap TT feedback for the Taylor--Green vortex over
80 steps. Higher ranks recover the uncompressed LKS velocity, while
rank four damps the decaying vortex amplitude. The profile comparison
also includes the analytical continuum velocity.}
\label{fig:TaylorGreenTT}
\end{figure}

The rank-four result shows why distribution accuracy and feedback accuracy must be checked separately. Its final distribution error is below one percent ($E_{f,\mathrm{TT}}^{80}=9.47\times10^{-3}$), yet its relative velocity error reaches $E_{u,\mathrm{TT}}^{80}=1.000$ and the final vortex profile is nearly lost (Figure~\ref{fig:TaylorGreenTT}). At low Mach number, the dominant density background can be approximated accurately while the smaller differences between opposite-direction populations, which carry momentum, are poorly resolved. This interpretation is consistent with the sensitivity estimate in Section~\ref{subsec:TTfeedbackError} and moment behavior in low-rank kinetic approximations \cite{coughlin2024robust,dektor2026interpolatory,asgrim2026tensor}. The equilibrium distribution itself contributes to these momentum-carrying components. All runs retain positive density and coefficients, so positivity alone is insufficient to select a rank that preserves the velocity. The velocity criterion in \eqref{eq:LKSoutputRequirements} must be checked explicitly.

Exact overlaps separate these compression effects from measurement noise. We next keep the rank and sweep count fixed and vary readout precision to examine how the reconstructed trajectory approaches the exact-overlap baseline as sampling error decreases.

\FloatBarrier
\subsection{Simulated Hadamard readout and measurement cost}
\label{subsec:TTnumericalExample}

The four-step D2Q9 test varies the precision of the core readout in Section~\ref{subsec:TTreadout} while holding the reconstruction ranks and sweep count fixed. It measures how sampling error changes the nonlinear trajectory and the prescribed measurement budget in \eqref{eq:TThybridFinalTotalCost}. We use an $8\times8$ periodic grid with $\Delta x=\Delta t=1$, $c_s^2=1/3$, $A_{\mathrm{LKS}}=0.3$, and $N_t=4$, for which Equation~\eqref{eq:LKSviscosity} gives nominal viscosity $\nu=0.1$. The initial fields are
\[
 \varrho^0=1,\qquad
 u_x^0=0.04\sin(2\pi x/8)\cos(2\pi y/8),\qquad
 u_y^0=-0.04\cos(2\pi x/8)\sin(2\pi y/8).
\]
We initialize the distributions from the quadratic equilibrium without
the gradient correction,
\[
 f_i^0=w_i\left[1+3\bb c_i\cdot\bb u^0
       +\frac92(\bb c_i\cdot\bb u^0)^2-\frac32|\bb u^0|^2\right],
\]
which reproduces these density and velocity fields. Starting with the
first collision, we apply the full gradient-corrected equilibrium
\eqref{eq:LKSequilibriumExplicit} with periodic centered differences,
taking as the reference the direct double-precision evolution of this same
nonlinear discrete recurrence from the uncompressed initial data.

We tensorize the distribution in the order $(i,x_1,x_2,x_3,y_1,y_2,y_3)$
from most to least significant coordinate bit, yielding
$p=7$ modes of sizes $(9,2,2,2,2,2,2)$.
Initial TT-SVD with rank cap six produces ranks $(6,6,6,6,4,2)$ with
relative initial distribution error $5.42\times10^{-4}$, storing
338 core entries compared with 576 grid entries (at most 72 entries in the largest local core). Holding these ranks fixed, we perform
two alternating orthogonalized-frame core sweeps per time step, first left to right and then
right to left; this fixed-budget protocol records the achieved fitting residual instead of
assuming that two sweeps meet a preassigned target tolerance.

We compare exact and sampled overlaps, recomputing the coefficients from the fitted distribution after every step. Dense arrays supply the moments, coefficients, scaling, and ideal overlaps; TT cores store the candidate and its alternating updates. The exact-overlap trajectory retains initial compression and finite-sweep fitting errors and serves as the baseline for the added sampling error. Quantum gate costs and the costs of TT coefficient evaluation and reversible coefficient access enter the analytical resource model; they are not measured by this statistical test.

For sampled updates, we allocate
\[
 \eta_{k,\ell,s}
 =\eta_{\mathrm{rel}}\frac{\|\bb f_{\mathrm{TT}}^k\|}{\sqrt p},
 \qquad \eta_{\mathrm{rel}}\in\{3\times10^{-3},10^{-3}\},
 \qquad \delta_{k,\ell,s}=\frac{0.05}{4\cdot2\cdot7}.
\]
Using the shot count prescribed by \eqref{eq:TTshotsPerEntry} for each overlap with ideal expectation $\mu$, we draw
the aggregate count
\[
 Z\sim\operatorname{Binomial}\left(N_{\mathrm{meas}},\frac{1+\mu}{2}\right),
 \qquad \widehat\mu=2Z/N_{\mathrm{meas}}-1.
\]
This binomial draw reproduces the exact law of the empirical Hadamard-test mean
without individually simulating the indicated number of shots. Ten
independent random seeds, 17--26, are used for each sampling tolerance, while fitting residuals are evaluated directly without extra sampling and circuit and scaling errors are excluded.

At every step, independent evaluation of the weighted collision, weighted streaming, and amplitude-recovery sequence matches the direct collision--streaming
target to a relative discrepancy below $2.9\times10^{-16}$.
Across these runs, all evaluated densities exceed $0.9985$ and equilibrium coefficients exceed
$0.02457$. Rather than relying on a nodewise sum of one, we retain the actual coefficient-sum bound
$s_{\max}^k$: with
$\eta_{\mathrm{rel}}=10^{-3}$, its final-update mean reaches
$1.0000480$, showing why discrete divergence perturbations in the gradient-corrected
coefficients must be checked even for initially divergence-free data.

\begin{figure}[htbp]
\centering
\subfigure[Distribution error]{
 \includegraphics[width=0.47\linewidth]{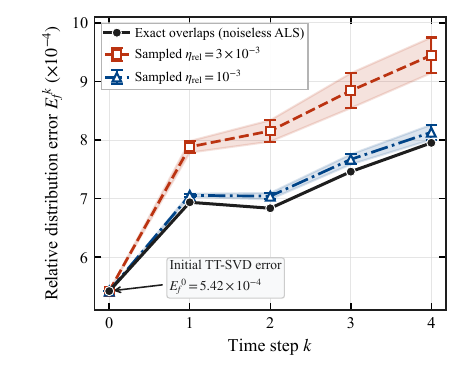}
}
\hfill
\subfigure[Mean shot budget per step]{
 \includegraphics[width=0.47\linewidth]{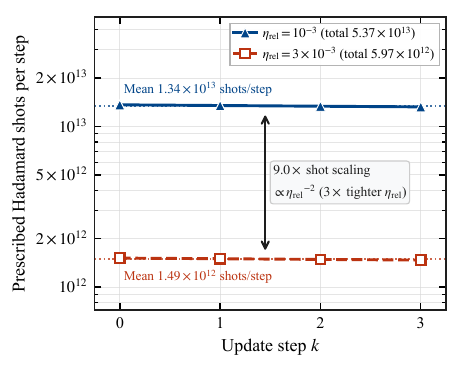}
}
\caption{Classical simulation of ideal TT core readout in the nonlinear
D2Q9 example. Error bars show one sample standard deviation.
The ranks and sweep counts are fixed; coefficients and measurement
budgets vary with time.}
\label{fig:TTfeedback}
\end{figure}

Let $\bb f_{\mathrm{ref}}^k$ and $\bb u_{\mathrm{ref}}^k$ denote the
direct discrete reference and define
\[
 E_f^k=\frac{\|\bb f_{\mathrm{TT}}^k-\bb f_{\mathrm{ref}}^k\|}
                  {\|\bb f_{\mathrm{ref}}^k\|},\qquad
 E_u^k=\frac{\|\bb u_{\mathrm{TT}}^k-\bb u_{\mathrm{ref}}^k\|}
                  {\|\bb u_{\mathrm{ref}}^k\|}.
\]
Tightening the per-core sampling tolerance drives the reconstructed trajectory toward the exact-overlap baseline, which has no sampling error while retaining initial compression and finite-rank fitting errors (Table~\ref{tab:TTfeedback} and Figure~\ref{fig:TTfeedback}, where sampled entries report means and sample standard deviations over the ten runs). Lowering $\eta_{\mathrm{rel}}$ from $3\times10^{-3}$ to $10^{-3}$ reduces the mean final relative fitting residual $\|\bb f_{\mathrm{TT}}^4-\bb y^3\|/\|\bb y^3\|$ from $5.15\times10^{-4}$ to $3.08\times10^{-4}$, approaching the exact-overlap residual of $2.69\times10^{-4}$.

\begin{table}[htbp]
\centering\small
\caption{Nonlinear TT feedback after four steps.
Uncertainties are sample standard deviations over ten runs.}
\label{tab:TTfeedback}
\renewcommand{\arraystretch}{1.2}
\setlength{\tabcolsep}{5pt}
\begin{tabular*}{\linewidth}{@{\extracolsep{\fill}}lccc@{}}
\toprule
Core readout & \makecell{$E_f^4$\\$(\times10^{-4})$}
 & \makecell{$E_u^4$\\$(\times10^{-2})$}
 & \makecell{Mean total shots\\$(\times10^{12})$} \\
\midrule
Exact overlaps & $7.95$ & $1.65$ & --- \\
$\eta_{\mathrm{rel}}=3\times10^{-3}$
 & $9.45\pm0.30$ & $2.77\pm0.14$ & $5.97$ \\
$\eta_{\mathrm{rel}}=10^{-3}$
 & $8.13\pm0.13$ & $1.847\pm0.045$ & $53.7$ \\
\bottomrule
\end{tabular*}
\par\smallskip
{\footnotesize\raggedright
Shot counts are prescribed budgets represented by binomial sampling,
not executed hardware shots. Exact overlaps use classical contractions.\par}
\end{table}

The sufficient shot allocation depends on the normalization of each overlap, the core sizes, and the local error budgets in Section~\ref{sec:resources}. For the tighter tolerance, a representative overlap requires precision approximately $3.2\times10^{-5}$. Summing the allocations over cores, sweeps, and time steps gives the mean total budget $5.37\times10^{13}$ in Table~\ref{tab:TTfeedback}. The amplitude scale and core tolerances change during the trajectory, so this sum differs from an extrapolation based on the first step, as illustrated in Appendix~\ref{app:readoutBudget}. Binomial sampling represents these prescribed repetitions without executing individual hardware shots; the allocation is a sufficient budget, not a sampling lower bound.

Reducing $\eta_{\mathrm{rel}}$ by a factor of three increases the prescribed shot budget approximately ninefold, from $5.97\times10^{12}$ to $5.37\times10^{13}$. The mean final distribution error decreases from $9.45\times10^{-4}$ to $8.13\times10^{-4}$, approaching the exact-overlap value $7.95\times10^{-4}$. The velocity comparison gives complementary information: its mean relative error decreases from $2.77\times10^{-2}$ to $1.847\times10^{-2}$, toward the exact-overlap value $1.65\times10^{-2}$. Thus tighter sampling improves the feedback quantity even when the change in distribution error is small, but both metrics approach a trajectory that still contains initial compression and finite-sweep fitting errors.

These results couple the choice of readout precision to the reconstruction budget. Once sampling error is small compared with the error remaining at the chosen rank and sweep count, further sampling cannot remove that remaining error. Ranks, sweep counts, and sampling tolerances must therefore be assessed together using the required physical outputs. The shot budgets count repetitions; total gate counts also include the input preparation, frozen-step evolution, and test-state circuits in \eqref{eq:TThybridMeasurementCost}.

\FloatBarrier
\section{Conclusion}\label{sec:conclusion}

We have constructed a TT interface that makes the output of a quantum lattice kinetic step available for the next nonlinear flow update. The interface retains the distribution at its physical amplitude scale in classical cores. These cores supply the local moments and gradients for coefficient evaluation and the state used to prepare the next quantum input. Within a step, energy scaling permits the frozen-coefficient construction, coordinate recovery returns the output to physical variables, and signed overlap estimates multiplied by the known normalization factor produce candidate output cores. Residual and admissibility checks determine whether to accept them. Storing the accepted cores removes the need to replay earlier quantum steps during subsequent readout.

The analysis distinguishes the conditions for encoding one step from those for controlling feedback over many steps. The circuit retains the actual nodewise coefficient sum, including the correction caused by discrete velocity divergence. The multi-step error bound uses positive, unit-sum reference coefficients, the stated density bounds, and Lipschitz sensitivity of the equilibrium map. Under these assumptions, local coefficient and update defects propagate to distribution and macroscopic feedback errors. The resource model then counts classical TT operations, circuit preparation and execution, sampling, and verification at the required precision. Bounded TT ranks reduce stored parameters to $\mathcal{O}(pn_{\max}r^2)$, while independent Hadamard tests retain an $\mathcal{O}(\epsilon_\mu^{-2})$ dependence on overlap precision.

The numerical diagnostics identify why compression, feedback accuracy, and readout precision require separate control. Classical Taylor--Green refinement shows nearly second-order velocity convergence at fixed Reynolds number and decreasing Mach number, and verifies the single-step scaled-update algebra. A classical three-dimensional cavity snapshot demonstrates static QTT compression, while the periodic 80-step feedback test examines repeated reconstruction. In that test, rank four nearly eliminates the vortex even though the final distribution error is below $1\%$; higher ranks recover the velocity more accurately. The density background can dominate the distribution norm while concealing errors in the smaller momentum-carrying components. The four-step statistical readout test shows a different limitation: tighter overlap estimates improve both distribution and velocity errors, but the trajectory approaches an exact-overlap baseline that retains initial compression and finite-sweep fitting errors. Its prescribed sufficient sampling budget quantifies the precision cost of this approach, rather than a necessary lower bound on readout.

These results define concrete requirements for a complete implementation. Reversible coefficient access must be compiled from the classical cores and combined with rank and sweep selection checked through \eqref{eq:TTquantumFitResidual}. Independent Hadamard-test estimation is a candidate for replacement by coherent amplitude estimation when inverse preparation, inverse step circuits, and reflections are available; the corresponding $\mathcal{O}(\epsilon_\mu^{-1})$ query scaling must still be matched to reconstruction accuracy and circuit costs. Wall-bounded evolution additionally requires an encoded boundary operator and control of the mass drift observed in the cavity calculation. A full implementation can then be compared with purely classical TT propagation at matched output tolerances \eqref{eq:LKSoutputRequirements} and confidence. The present classical and statistical diagnostics provide the interface tests and error accounting for that comparison; quantum runtime advantage remains to be established.

\section*{CRediT authorship contribution statement}
\textbf{Yongwei Guo:} Software, Validation (supporting), Visualization, Writing -- review \& editing.
\textbf{Yuan Yu:} Conceptualization, Methodology, Formal analysis, Supervision, Funding acquisition, Writing -- review \& editing.
\textbf{Yue Yu:} Methodology (lead), Formal analysis (lead), Validation (lead), Funding acquisition, Writing -- review \& editing.
\textbf{Yuting Zhou:} Software, Validation (supporting), Visualization, Writing -- review \& editing.

\section*{Declaration of competing interest}
The authors declare that they have no known competing financial interests or personal relationships that could have appeared to influence the work reported in this paper.

\section*{Data availability}
Data and source code supporting the findings of this study are available from the corresponding author upon reasonable request.

\section*{Acknowledgments}
Yuan Yu was supported by the National Natural Science Foundation of China (No.\ 12101527) and the Natural Science Foundation of Hunan Province (No.\ 2026JJ60005).
Yue Yu was supported by NSFC grant (No.\ 12301561), the Key Project of Scientific Research Project of Hunan Provincial Department of Education (No.\ 24A0100), the Science and Technology Innovation Program of Hunan Province (No.\ 2025RC3150), the general program of Hunan Provincial Natural Science Foundation (No.\ 2026JJ50003), and partially supported by NSFC grant No.\ 12341104.
This research was supported in part by the Science and Technology Innovation Program of Hunan Province (No.\ 2026RC3172), the 111 Project (No.\ D23017), and Program for Science and Technology Innovative Research Team in Higher Educational Institutions of Hunan Province of China. Computational resources were provided by the High Performance Computing Platform of Xiangtan University.

\appendix
\renewcommand{\thesection}{\Alph{section}}

\section{Energy-scaled propagation for prescribed coefficients}\label{app:linearBE}

The single-step construction uses energy-scaled propagation with coefficients held fixed during that step. Sections~\ref{subsec:structuredBlockEncoding} and~\ref{subsec:prescribedTrajectory} recall the corresponding linear results from the companion study~\cite{HeYu2026ADE} and specialize them to a prescribed LKS coefficient trajectory. Coherent propagation along such a trajectory supplies background for the quantum operator; the nonlinear algorithm instead reconstructs classical cores and recomputes coefficients after each step. Section~\ref{subsec:appCompressedPrescribed} develops the auxiliary compressed recurrence for prescribed coefficients used in the recovery and normalization argument of Proposition~\ref{prop:TTnonlinear}.

\subsection{Compact recurrence and energy-scaled block-encoding preliminaries}\label{subsec:structuredBlockEncoding}

To make the stepwise block-encoding hypotheses explicit, we recall the compact prescribed-coefficient recurrence under positive, nodewise-normalized equilibrium weights.

On a periodic lattice with $q$ discrete velocities and $N$ spatial nodes, when the equilibrium coefficients $a_i(t_k,\bb x_j)$ are prescribed and collision couples distributions through the zeroth moment $\bb\varrho^k=E_I\bb f^k$, combining BGK collision with permutation streaming $P=\diag(P_0,\ldots,P_{q-1})$ yields the compact recurrence
\[
 \bb f^{k+1}=M^k\bb f^k,\qquad
 M^k=P\bigl(\alpha I_{qN}+\beta\mathcal A^k E_I^\top E_I\bigr),
\]
where $\alpha=1-1/\tau^*$, $\beta=1/\tau^*$, $\mathcal A^k=\diag(A_0^k,\ldots,A_{q-1}^k)$, and $A_i^k=\diag(a_i(t_k,\bb x_0),\allowbreak\ldots,\allowbreak a_i(t_k,\bb x_{N-1}))$. At full relaxation $\tau^*=1$ ($\alpha=0,\beta=1$), this reduces to $M^k=P\mathcal A^k E_I^\top E_I$.

We assume there exists a grid-independent constant $a_*>0$ such that the equilibrium coefficients satisfy the uniform positivity condition
\begin{equation}\label{eq:positivity}
 a_i(t,\bb x_j)\geq a_*,\qquad i=0,\ldots,q-1,\quad j\in[N],
\end{equation}
together with the nodewise normalization $\sum_{i=0}^{q-1}a_i(t_k,\bb x_j)=1$. With the energy-scaled distribution vector $\bb g^k=(\mathcal A^k)^{-1/2}\bb f^k$, the two-time-level scaled recurrence reads
\begin{equation}\label{gtime}
 \bb g^{k+1}=B^k\bb g^k,\qquad
 B^k=(\mathcal A^{k+1})^{-1/2}M^k(\mathcal A^k)^{1/2}
    =\widetilde P^k\widetilde C^k,
\end{equation}
where the weighted streaming and symmetrized collision factors are
\begin{gather*}
 \widetilde P^k=(\mathcal A^{k+1})^{-1/2}P(\mathcal A^k)^{1/2},\qquad
 \widetilde C^k=\Pi^k+\alpha(I_{qN}-\Pi^k),\\
 \Pi^k=Q^k(Q^k)^\dagger,\qquad Q^k=(\mathcal A^k)^{1/2}E_I^\top.
\end{gather*}

\begin{theorem}[\cite{HeYu2026ADE}]\label{thm:norm}
 Let $P_i\ket j=\ket{\pi_i(j)}$, where $\pi_i$ is a permutation. Under \eqref{eq:positivity}, $\sum_{i=0}^{q-1}A_i^k=I_N$, and $\tau^*\geq1/2$, for each time step $k$ we have $(Q^k)^\dagger Q^k=I_N$, so $\Pi^k$ is an orthogonal projection and
 \[
 \|\widetilde C^k\|=1,\qquad
 \|B^k\|\leq\|\widetilde P^k\|=\rho^k,
 \]
 where the one-step weighted-streaming ratio is
 \begin{equation}\label{eq:coefficientRatioBound}
 \rho^k=\max_{i,j}
 \sqrt{\frac{a_i(t_k,\bb x_j)}{a_i(t_{k+1},\bb x_{\pi_i(j)})}}.
 \end{equation}
\end{theorem}

\begin{lemma}[Growth bound for time-dependent coefficients~\cite{HeYu2026ADE}]\label{lem:spaceTimeGrowth}
 Under the hypotheses of Theorem~\ref{thm:norm}, let streaming be periodic and let $a_i$ be positive and continuously differentiable on $[0,T]\times\Omega$. Suppose there exists a grid-independent constant $\ell_{\mathrm{st}}\geq0$ such that
 \begin{equation}\label{eq:spaceTimeSmoothness}
  \max_i\sup_{(t,\bb x)\in[0,T]\times\Omega}
  \left|(\partial_t+\bb c_i\cdot\nabla)\log a_i(t,\bb x)\right|
  \leq\ell_{\mathrm{st}}.
 \end{equation}
 Then $\|B^k\|\leq\rho^k\leq\exp(\ell_{\mathrm{st}}\Delta t/2)$ for $0\leq k<N_t$, and $\|B^{s-1}\cdots B^r\|\leq\exp\bigl(\ell_{\mathrm{st}}(t_s-t_r)/2\bigr)$ for $0\leq r<s\leq N_t$.
\end{lemma}

With the local state-preparation unitary $V^k\ket{0^{\lceil\log_2 q\rceil}}\ket j=\sum_{i=0}^{q-1}\sqrt{a_i(t_k,\bb x_j)}\ket i\ket j=Q^k\ket j$, the operator $R_\Pi^k=2\Pi^k-I$ forms a unitary reflection and gives a $(1,1,0)$-block-encoding of $\widetilde C^k=\frac{1+\alpha}{2}I+\frac{1-\alpha}{2}R_\Pi^k$ (which reduces to $\Pi^k$ at full relaxation $\tau^*=1$). Composing this collision encoding with controlled-rotation weighted streaming yields the following block-encoding bound.

\begin{proposition}[\cite{HeYu2026ADE}]\label{prop:structuredBE}
 Under coherent coefficient-oracle and permutation-streaming access, let $\Lambda\geq\max\{1,\rho^0,\ldots,\rho^{N_t-1}\}$. Then each $B^k$ admits a $(\Lambda,2,0)$-block-encoding $U_B^k$ in the exact-oracle model using $\mathcal O(1)$ oracle calls, and an $\eta$-accurate approximation to $B^k/\Lambda$ in operator norm using $\mathcal O(1)$ calls at $\mathcal O(\log(1/\eta))$-bit precision.
\end{proposition}

We define the cumulative normalization factor, scaled-norm ratio, and combined amplification parameter by
\begin{equation}\label{eq:quantumGrowth}
 \Gamma_T=\Lambda^{N_t},\qquad
 R_T=\frac{\|\bb g^0\|}{\|\bb g^{N_t}\|},\qquad K_T=\Gamma_T R_T.
\end{equation}

\begin{theorem}[Coherent time-marching complexity for prescribed coefficients~\cite{HeYu2026ADE}]\label{thm:tmComplexity}
 Under \eqref{eq:positivity} with fixed $a_*>0$, $\tau^*\geq1/2$, and the access assumptions of Proposition~\ref{prop:structuredBE}, assume an initial-state preparation unitary $U_0$ for $\ket{g^0}$ and its inverse are available and a constant-factor upper bound on $K_T$ is known. For $0<\varepsilon<1$, dilating unitarization with a logarithmic-size failure counter, terminal recovery by $(\mathcal A^{N_t})^{1/2}$, and amplitude amplification prepare a state $\ket{\widetilde f^{N_t}}$ satisfying $\|\ket{\widetilde f^{N_t}}-\ket{f^{N_t}}\|\leq\varepsilon$ with constant success probability using $\mathcal O(N_tK_T)$ evolution-oracle queries, with single-step block-encoding precision $\eta=\mathcal O(\varepsilon\sqrt{a_*}/(N_tK_T))$.
\end{theorem}

In the nonlinear incompressible LKS considered in the main text, the coefficients depend on the evolving velocity and its gradient, and the gradient correction can perturb the unit-sum identity $\sum_i a_i[\bb u^k](\bb x_j)=1$; we adapt the weighted-coordinate framework accordingly within each frozen-coefficient quantum step.

\subsection{Evolution with a prescribed coefficient trajectory}\label{subsec:prescribedTrajectory}

\begin{proposition}[Quantum evolution for a prescribed LKS coefficient trajectory]
\label{prop:LKSextension}
 Consider \eqref{eq:LKSfrozen} on a periodic lattice with fixed $q$.
 Assume \eqref{eq:LKScoefficientConditions}, coherent time-indexed
 coefficient access, and the initial-state and amplification assumptions
 of Theorem~\ref{thm:tmComplexity}. Choose
\[
 \Lambda\geq\max\left\{1,
 \max_{k,i,j}\sqrt{\frac{a_i[\overline{\bb u}^{\,k}](\bb x_j)}
                         {a_i[\overline{\bb u}^{\,k+1}](\bb x_{\pi_i(j)})}}\right\}.
\]
 With $\bb g^k=(\mathcal A^k)^{-1/2}\bb f^k$, define $\Gamma_T$, $R_T$,
 and $K_T$ by \eqref{eq:quantumGrowth}. Then the normalized distribution
 state of this linear subproblem can be prepared to error $\varepsilon$
 with constant success probability using $\mathcal O(N_tK_T)$
 evolution-oracle queries for coherent time marching.
 Constants may depend on $q$ and the fixed positivity margin $a_*$.
\end{proposition}
\begin{proof}
 Set $Q^k=(\mathcal A^k)^{1/2}E_I^\top$. The moment condition gives
 $(Q^k)^\dagger Q^k=I$, so $\Pi^k=Q^k(Q^k)^\dagger$ is an orthogonal
 projection. The scaled step is
\[
 B^k=(\mathcal A^{k+1})^{-1/2}P(\mathcal A^k)^{1/2}\Pi^k.
\]
 The first factor has norm bounded by $\Lambda$, and the projection has
 a normalization-one block-encoding. Local preparation of the $q$
 amplitudes $\sqrt{a_i[\overline{\bb u}^{\,k}]}$ has constant query cost for fixed $q$ under
 the stated access model. Consequently, Proposition~\ref{prop:structuredBE}
 applies with full relaxation. Recovery by $(\mathcal A^{N_t})^{1/2}$
 has success probability at least $a_*$. The proof of
 Theorem~\ref{thm:tmComplexity} then applies without further changes.
\end{proof}

If the prescribed coefficients admit positive smooth interpolants
satisfying \eqref{eq:spaceTimeSmoothness}, Lemma~\ref{lem:spaceTimeGrowth}
again gives $\Gamma_T\leq\exp(\ell_{\mathrm{st}}T/2)$ for the corresponding
choice of $\Lambda$. For gradient-dependent coefficients, verifying this
condition typically requires controlling derivatives of the velocity
gradients as well as derivatives of the velocity field, so this smoothness hypothesis must be checked for the
chosen discretization and refinement regime.

\subsection{Compressed energy-scaled recurrence for prescribed coefficients}\label{subsec:appCompressedPrescribed}

To isolate how local compression and operator defects propagate along a prescribed, positive, normalized coefficient trajectory before nonlinear feedback is introduced, take $\Lambda\geq\max\{1,\max_k\|B^k\|\}$ and recall $\Gamma_T=\Lambda^{N_t}$, $R_T=\|\bb g^0\|/\|\bb g^{N_t}\|$, and $K_T=\Gamma_T R_T$ from \eqref{eq:quantumGrowth}. For prescribed coefficients, an auxiliary compressed recurrence propagates the scaled distribution according to
\begin{equation}\label{eq:TTmarching}
 \bb g_{\mathrm{TT}}^{k+1}
 =\mathcal R_{\delta_k}
       (\widehat B^k\bb g_{\mathrm{TT}}^{k}),
 \qquad k=0,\ldots,N_t-1.
\end{equation}
Here $\widehat B^k$ is the implemented operator, possibly obtained by
applying and rounding the collision and streaming factors separately.
More generally, all intermediate approximations can be included in the
one-step defect
\[
 \bb e^k=\bb g_{\mathrm{TT}}^{k+1}-B^k\bb g_{\mathrm{TT}}^{k}.
\]
If \eqref{eq:TTmarching} is used with
$\|\widehat B^k-B^k\|\leq\zeta_k$, then
\begin{equation}\label{eq:TTlocalDefect}
 \|\bb e^k\|\leq\delta_k+
                   \zeta_k\|\bb g_{\mathrm{TT}}^{k}\|.
\end{equation}
These local vector defects are measured in absolute Euclidean norm, distinguishing them from the normalized
block-encoding operator errors used in the quantum query analysis.

\begin{proposition}[Stability of compressed time marching]\label{prop:TTmarching}
 Under the assumptions of Theorem~\ref{thm:norm}, suppose
 $\|B^k\|\leq\Lambda$ and $\|\bb e^k\|\leq d_k$.
 Let $d_{\mathrm{init}}=\|\bb g_{\mathrm{TT}}^{0}-\bb g^0\|$ and define
\begin{equation}\label{eq:TTerrorBudget}
 E_T=\Gamma_T d_{\mathrm{init}}
       +\sum_{k=0}^{N_t-1}\Lambda^{N_t-1-k}d_k.
\end{equation}
 Then $\|\bb g_{\mathrm{TT}}^{N_t}-\bb g^{N_t}\|\leq E_T$.
 If the recovered vector satisfies
\[
 \bb f_{\mathrm{TT}}^{N_t}
  =(\mathcal A^{N_t})^{1/2}\bb g_{\mathrm{TT}}^{N_t}
      +\bb e_{\mathrm{rec}},\qquad
 \|\bb e_{\mathrm{rec}}\|\leq d_{\mathrm{rec}},
\]
 then $\|\bb f_{\mathrm{TT}}^{N_t}-\bb f^{N_t}\|
 \leq E_T+d_{\mathrm{rec}}$. Whenever
 $E_T+d_{\mathrm{rec}}<\|\bb f^{N_t}\|$, the normalized vectors obey
\begin{equation}\label{eq:TTnormalizedError}
 \left\|\frac{\bb f_{\mathrm{TT}}^{N_t}}{\|\bb f_{\mathrm{TT}}^{N_t}\|}
             -\ket{f^{N_t}}\right\|
 \leq\frac{2(E_T+d_{\mathrm{rec}})}{\|\bb f^{N_t}\|}.
\end{equation}
\end{proposition}
\begin{proof}
 Subtracting the exact recurrence gives
\[
 \bb g_{\mathrm{TT}}^{k+1}-\bb g^{k+1}
 =B^k(\bb g_{\mathrm{TT}}^{k}-\bb g^k)+\bb e^k.
\]
 Iteration and the ordered-product bounds yield
 \eqref{eq:TTerrorBudget}. Since
 $\|(\mathcal A^{N_t})^{1/2}\|\leq1$, recovery does not increase
 the propagated absolute error. The final statement follows from
 the normalization inequality used in Theorem~\ref{thm:tmComplexity}.
\end{proof}

For example, with exact initialization, exact step application, and exact
recovery, it suffices to impose
\[
 \delta_k\leq
 \frac{\varepsilon\sqrt{a_*}\|\bb g^0\|}{2N_tK_T}
 \quad\text{for all }k
\]
to obtain final normalized error at most $\varepsilon<1$. Indeed,
$E_T\leq N_t\Gamma_T\max_k\delta_k
\leq\varepsilon\sqrt{a_*}\|\bb g^{N_t}\|/2
\leq\varepsilon\|\bb f^{N_t}\|/2$.
With approximate operators and recovery, the same budget is divided
among the defects in \eqref{eq:TTlocalDefect} and $d_{\mathrm{rec}}$.
For $\Lambda=1$, the bound reduces to the sum of local defects.
For smooth prescribed coefficients, the interval bounds of
Lemma~\ref{lem:spaceTimeGrowth} may replace the powers of $\Lambda$.

This auxiliary procedure uses the same scaled recurrence as the prescribed-coefficient
coherent propagation in Sections~\ref{subsec:structuredBlockEncoding} and~\ref{subsec:prescribedTrajectory}, yet performs classical
tensor contractions and rounding rather than the physical-scale feedback
loop of Section~\ref{sec:TThybrid}.
Because propagation is purely classical, it requires neither a failure counter nor postselection or amplitude
amplification, and its work is governed by TT ranks, mode sizes, and
rounding tolerances rather than by the quantum query bound $\mathcal O(N_tK_T)$. For an MPO of maximum rank $R$
and a state of maximum rank $r$, direct operator application produces intermediate
state ranks at most $Rr$ prior to rounding. Consequently, these intermediate rank bounds together with coefficient-construction and truncation costs must all be recorded, whereas only the current state needs to be retained unless a full trajectory history is requested.

\section{Tensorization and tensor-train algebra}\label{subsec:TTrepresentation}

The operations in Section~\ref{subsec:TTworkingState} act on TT cores whose index ordering and intermediate ranks determine storage and work. This appendix specifies that tensorization, the vector and matrix cores, and the contraction and rounding rules used in the feedback calculation.

\subsection{Vector tensorization, cores, and ranks}

The tensor-train (TT) format represents a multidimensional array by a
chain of smaller arrays called cores~\cite{oseledets2011TT}.
Tensorization reshapes a state vector into a higher-order array without changing its entries, so tensor order need not equal spatial dimension; low-rank compression then restricts the auxiliary dimensions connecting adjacent cores.

In our application, the TT parameters are classical arrays representing
the \emph{physical distribution} $\bb f^k$. We fix a bijection between $(i,j)$ and tensor indices
$(\xi_1,\ldots,\xi_p)$, with mode sizes $n_s$ satisfying
$\prod_s n_s=qN$. For example, on a one-dimensional grid with
$N=2^m$ nodes, we write the node index in binary as
\[
 j=\sum_{\ell=1}^{m}2^{m-\ell}b_\ell,
 \qquad b_\ell\in\{0,1\},\qquad 0\leq j<N.
\]
Taking the direction index as the first mode and the binary digits as
the remaining modes gives
\[
 (i,j)\longleftrightarrow(\xi_1,\ldots,\xi_{m+1})
 =(i,b_1,\ldots,b_m),\qquad
 n_1=q,\quad n_2=\cdots=n_{m+1}=2.
\]
With this zero-based indexing, the tensorized distribution
$\mathsf T^k$ is defined entrywise by
\begin{equation}\label{eq:TTbinaryTensorization}
 \begin{aligned}
 \mathsf T^k(i,b_1,\ldots,b_m)
 &:=f_{i,\,\sum_{\ell=1}^{m}2^{m-\ell}b_\ell}^{k},\\
 \mathsf T^k&\in\mathbb R^{q\times\underbrace{2\times\cdots\times2}_{m\text{ binary modes}}},
 \qquad p=m+1.
 \end{aligned}
\end{equation}
For $m=3$ and $j=5=(101)_2$, this mapping gives $\mathsf T^k(i,1,0,1)=f_{i,5}^k$, showing that $\mathsf T^k$ simply re-indexes the full distribution prior to
compression. On multi-axis grids, we apply this splitting construction to each Cartesian coordinate under a fixed mode ordering, because the chosen index order directly affects attainable TT ranks.

At time level $k$, we store the classical TT cores
\[
 \mathsf F_s^k\in\mathbb R^{r_{s-1}^k\times n_s\times r_s^k},
 \qquad s=1,\ldots,p,\qquad r_0^k=r_p^k=1.
\]
Contracting this core chain via \eqref{eq:TTrepresentation} reconstructs the unnormalized physical-distribution approximation directly at physical scale without forming the dense lattice array.

Each core has one physical index $\xi_s$ and two auxiliary indices
of sizes $r_{s-1}^k$ and $r_s^k$. Fixing $\xi_s$ selects the matrix
slice $\mathsf F_s^k(\xi_s)$. The symbol $\Join$ denotes the TT
rank-core product: it contracts the shared auxiliary index of adjacent
cores while retaining their physical indices. Once all physical indices
are fixed, this contraction reduces to ordinary multiplication of the
selected matrix slices. The boundary ranks $r_0^k=r_p^k=1$ make the
result in \eqref{eq:TTrepresentation} a scalar. Equivalently, the full
tensorized distribution is represented by the core chain
$\mathsf F_1^k\Join\cdots\Join\mathsf F_p^k$.
For three modes, this contraction is explicitly
\[
 (\bb f_{\mathrm{TT}}^k)_{i,j}
 =\sum_{\alpha_1=1}^{r_1^k}\sum_{\alpha_2=1}^{r_2^k}
 \mathsf F_1^k(1,\xi_1,\alpha_1)
 \mathsf F_2^k(\alpha_1,\xi_2,\alpha_2)
 \mathsf F_3^k(\alpha_2,\xi_3,1).
\]
The sums contract shared auxiliary indices.

The integers $r_s^k$ are the TT ranks. They control the coupling between
the first $s$ modes and the remaining modes. More precisely, grouping
these two sets of indices into the row and column indices of a matrix
gives a matrix of rank at most $r_s^k$ for the represented tensor.
If all TT ranks are one, each entry is simply a product of scalar
functions of the separate indices. Larger ranks allow sums of such
coupled contributions. The cores contain
$\sum_s n_sr_{s-1}^kr_s^k$ numbers instead of $\prod_s n_s=qN$;
when $n_s\leq n_{\max}$ and $r_s^k\leq r$, this is at most
$p n_{\max}r^2$. The storage reduction therefore depends on the ranks
needed at the requested accuracy, not just on the number of modes.

The stored cores allow individual entries to be evaluated on demand without
expanding the full distribution, with initial cores constructed from classical
data and subsequent cores fitted from quantum overlaps.

Throughout our notation, the subscript $\mathrm{TT}$ marks computed low-rank fields and sans-serif letters
denote their cores, while superscripts $k$ and subscripts $s$ index physical time steps and core positions; because TT factorizations admit gauge freedom, orthogonalization can modify individual cores without changing the represented vector.

Computed moments, velocity, and coefficients are denoted by
$\bb\varrho_{\mathrm{TT}}^k$, $\bb m_{\mathrm{TT}}^k$,
$\bb u_{\mathrm{TT}}^k$, and $\bb a_{\mathrm{TT}}^k$, respectively.
The direction block $\bb a_{i,\mathrm{TT}}^k$ approximates the exact
functional $a_i[\bb u(\bb f_{\mathrm{TT}}^k)]$, including errors in
moment evaluation, division, derivatives, and compression. Coefficient
fields have their own cores, denoted $\mathsf A_s^k$ after tensorizing
$(i,j)$ in the same order. These three-index arrays are distinct from
the diagonal physical operator $\mathcal A^k$. In particular,
\[
 a_{i,\mathrm{TT}}^k(\bb x_j)
 =\mathsf A_1^k(\xi_1)\Join\cdots\Join\mathsf A_p^k(\xi_p).
\]
Here the bold vector $\bb a_{i,\mathrm{TT}}^k$ collects these nodewise scalar values (written without boldface at a specified grid node) across the spatial lattice, and the TT ranks of the coefficient fields need not equal those of the distribution.

\subsection{Matrix cores and operations in TT format}

A TT matrix represents a linear map between tensorized vectors.
For simplicity, let $H\in\mathbb R^{n\times n}$ with
$n=\prod_{s=1}^p n_s$, and use the same tensorization for its row and
column indices. Write these indices as
$\bb\xi=(\xi_1,\ldots,\xi_p)$ and
$\bb\zeta=(\zeta_1,\ldots,\zeta_p)$, respectively. The matrix TT format is
\[
 H_{\bb\xi,\bb\zeta}
 =\mathsf H_1(\xi_1,\zeta_1)\Join\cdots
   \Join\mathsf H_p(\xi_p,\zeta_p),\qquad
 \mathsf H_s\in\mathbb R^{R_{s-1}\times n_s\times n_s\times R_s},
 \quad R_0=R_p=1.
\]
The two middle indices of each core select an output coordinate
$\xi_s$ and an input coordinate $\zeta_s$. Fixing them leaves a matrix
$\mathsf H_s(\xi_s,\zeta_s)\in\mathbb R^{R_{s-1}\times R_s}$;
the operator $\Join$ contracts adjacent auxiliary indices, yielding one scalar
entry of $H$. Thus a matrix core has four indices, whereas a vector
core has three. This representation is also called a matrix product
operator (MPO).

To see how it acts on a vector, let $\bb v$ have TT cores
$\mathsf V_s(\zeta_s)\in\mathbb R^{r_{s-1}\times r_s}$.
The standard matrix--vector product sums over the input coordinates according to
\[
 (H\bb v)_{\bb\xi}
 =\sum_{\zeta_1,\ldots,\zeta_p}
       H_{\bb\xi,\bb\zeta}v_{\bb\zeta}.
\]
Performing these summations locally on each core gives, prior to compression,
the output $\bb w=H\bb v$ with core slices
\[
 \mathsf W_s(\xi_s)
 =\sum_{\zeta_s=1}^{n_s}
     \mathsf H_s(\xi_s,\zeta_s)\otimes\mathsf V_s(\zeta_s)
 \in\mathbb R^{(R_{s-1}r_{s-1})\times(R_sr_s)}.
\]
Here $\otimes$ forms the Kronecker product of two local core slices while
$\Join$ contracts successive cores along the TT chain, so contracting the slices
$\mathsf W_s$ yields $H\bb v$ without expanding either the full matrix
or the full vector. In the present algorithm, TT matrices represent
coordinate shifts and discrete difference operators used to evaluate
$\nabla_h\bb u_{\mathrm{TT}}^k$.

Pointwise multiplication is different: it pairs entries at the same
grid index and involves no sum over an input coordinate. If $\bb a$
and $\bb v$ have compatible tensorizations and cores $\mathsf A_s$
and $\mathsf V_s$, then $\bb w=\bb a\odot\bb v$ has slices
\[
 \mathsf W_s(\xi_s)=\mathsf A_s(\xi_s)\otimes\mathsf V_s(\xi_s).
\]
This operation forms velocity from momentum and reciprocal density,
evaluates products in the equilibrium coefficients, and applies
diagonal energy scalings. Equivalently, $\diag(\bb a)$ has matrix
core slices $\delta_{\xi_s\zeta_s}\mathsf A_s(\xi_s)$, where
$\delta$ is the Kronecker delta.

Both operations multiply auxiliary bond sizes and can therefore
increase TT ranks; subsequent TT rounding compresses the resulting cores via orthogonalization and truncated singular value decompositions. We write $\mathcal R_\delta(\bb w)$
for a rounded vector satisfying
\[
 \|\mathcal R_\delta(\bb w)-\bb w\|\leq\delta.
\]
Imposing a fixed rank cap alone does not guarantee this certified tolerance. All algebraic and rounding steps in this appendix act classically on stored TT cores, whereas frozen-step quantum propagation and overlap-based reconstruction of updated cores are carried out in
Sections~\ref{subsec:TTquantumStep} and~\ref{subsec:TTreadout}.

\section{Worked sampling-budget calculation}\label{app:readoutBudget}

The four-step readout test in Section~\ref{subsec:TTnumericalExample} assigns a sufficient number of samples to each measured overlap. We expand that calculation here to show how a physical core tolerance becomes an overlap precision and then a total prescribed budget. The count follows the allocation rule in Section~\ref{sec:resources}; it is neither a sampling lower bound nor a hardware runtime estimate.

The large shot budget can be traced to the entrywise precision required
by the corewise reconstruction. For a core of size $d$, the relation
$(\bb\theta_s)_t=\mathcal N_k\mu_t$ converts a core error tolerance
$\eta_{k,\ell,s}$ into an overlap tolerance
$\epsilon_\mu=\eta_{k,\ell,s}/(\mathcal N_k\sqrt d)$ in the
small-error regime. At the first step of this example, for
$\eta_{\mathrm{rel}}=10^{-3}$ and a core with $d=72$, the values are
approximately
\[
 \mathcal N_0=5.578,\qquad
 \|\bb f_{\mathrm{TT}}^0\|\approx4.0,\qquad
 \eta_{0,\ell,s}\approx1.5\times10^{-3},\qquad
 \epsilon_\mu\approx3.2\times10^{-5}.
\]
With $\delta_{0,\ell,s}=0.05/56$, the per-entry bound \eqref{eq:TTshotsPerEntry}
therefore prescribes about $2.4\times10^{10}$ shots for just one
overlap. The complete count follows by summing over the actual core
sizes rather than assigning the largest size to every core; in the
small-error regime, the bound \eqref{eq:TTshotsPerEntry} gives the sufficient
budget for one core update as
\begin{equation}\label{eq:TTexampleShotBudget}
 N_{\mathrm{core}}^{k,\ell,s}
 =d_s^{k,\ell}\left\lceil
 \frac{2\mathcal N_k^2d_s^{k,\ell}}{\eta_{k,\ell,s}^2}
 \log\frac{2d_s^{k,\ell}}{\delta_{k,\ell,s}}
 \right\rceil.
\end{equation}
For the fixed ranks in this experiment, the seven core sizes are
\[
 (d_1,\ldots,d_7)=(54,72,72,72,48,16,4),
 \qquad \sum_{s=1}^7d_s=338.
\]
Using the first-step values
$\mathcal N_0\approx5.57799245$ and
$\eta_{0,\ell,s}=10^{-3}(3.99840365)/\sqrt7
\approx0.00151125$, with $\delta_{0,\ell,s}=0.05/56$, gives
\[
 \begin{aligned}
 N_{\mathrm{meas}}^{0,\ell,s,t}\big|_{d_s=72}
 &\approx2.3523\times10^{10},\\
 N_{\mathrm{core}}^{0,\ell,s}\big|_{d_s=72}
 &\approx72(2.3523\times10^{10})
 =1.6937\times10^{12}.
 \end{aligned}
\]
Because each physical time step performs two alternating sweeps, the first-step shot budget is
\[
 \begin{aligned}
 N_{\mathrm{shots}}^0
 &=2\sum_{s=1}^7 d_s
 \left\lceil
 \frac{2(5.57799245)^2d_s}{(0.00151125)^2}
 \log\frac{2d_s}{0.05/56}
 \right\rceil\\
 &\approx1.3630\times10^{13}.
 \end{aligned}
\]
Holding these first-step parameters fixed over four steps would give
$4N_{\mathrm{shots}}^0\approx5.4522\times10^{13}$ shots. This fixed-initial extrapolation provides only a
scale estimate, not a uniform upper bound for the nonlinear trajectory, because
the amplitude scale and the allocated absolute tolerance evolve with
$k$. The prescribed trajectory budget is instead
\[
 N_{\mathrm{shots}}^{\mathrm{total}}
 =\sum_{k=0}^{3}\sum_{\ell=1}^{2}\sum_{s=1}^{7}
 N_{\mathrm{core}}^{k,\ell,s}.
\]
Evaluating this sum with the updated parameters gives the mean
$5.37\times10^{13}$ reported in Table~\ref{tab:TTfeedback}.
The failure-budget allocation controls all local sampling errors
simultaneously with probability at least $0.95$; it does not by itself
guarantee a specified final distribution error after nonlinear feedback.
In particular, for a single core update the sufficient budget scales as
\[
 \mathcal O\!\left(
 \frac{\mathcal N_k^2d^2}{\eta_{k,\ell,s}^2}
                  \log\frac{2d}{\delta_{k,\ell,s}}\right).
\]
Here one factor of $d$ counts the core entries while the other arises from assigning the entrywise tolerance $\eta_{k,\ell,s}/\sqrt d$ across the Euclidean core norm; because the confidence level contributes only
a logarithmic factor, entrywise overlap readout and inverse-square precision dominate the measurement cost.

\begin{remark}[Shot budgets and the required precision]
The $10^{13}$ budget is consistent with the
$\mathcal O(\epsilon_\mu^{-2})$ law for independent Hadamard-test
sampling: the required overlap precision is approximately
$3.2\times10^{-5}$, and many overlaps must be estimated.
This is a sufficient budget for the chosen estimator, not a lower
bound. With coherent circuit access, its inverse, and the required
reflections, amplitude estimation can improve the per-overlap
precision dependence to $\mathcal O(\epsilon_\mu^{-1})$ quantum
queries at fixed success probability~\cite{brassard2000amplitude};
that improvement is not used here.
\end{remark}

\section*{Declaration of generative AI and AI-assisted technologies in the manuscript preparation process}
During the preparation of this manuscript, the authors used OpenAI Codex to assist with manuscript structure, wording, and literature verification. After using this tool, the authors reviewed and edited the content as needed and take full responsibility for the content of the published article.

\bibliographystyle{elsarticle-num}
\bibliography{RefsQLBM}

\end{document}